\documentclass[12pt,preprint]{aastex} 

\usepackage{url}
\usepackage{natbib}
\usepackage{booktabs}
\usepackage[colorlinks,
        hypertexnames=false,
        pagebackref=true,
        linkcolor=blue,          
    citecolor=green,        
    filecolor=magenta,      
    urlcolor=cyan                   
]{hyperref}                                 

\newcommand{\kmps}{\hbox{~\rm{$\mbox{km s}^{-1}$}}}

\newcommand{\jfm}{ {\it J. Fluid Mech.}}
\newcommand{\cpcf}{{\it Comments Plasma Phys. Controlled Fusion}}

\shorttitle{Helicity injection and Eruption Phenomena}
\shortauthors{Vemareddy et al}

\begin{document}
\title{On the injection of helicity by shearing motion of fluxes in relation to Flares and CMEs}
\author{P.~ Vemareddy$^1$, A.~ Ambastha$^1$, R.~ A.~ Maurya$^2$ and J.~ Chae$^2$}
\affil{$^1$Udaipur Solar Observatory, Physical Research Laboratory, Udaipur-313
001,India.} \affil{$^2$Astronomy Program, Department of Physics and Astronomy, Seoul
National University, Seoul 151-747, Korea} \email{vema@prl.res.in, ambastha@prl.res.in,
ramajor@astro.snu.ac.kr, jcchae@snu.ac.kr}

\begin{abstract}
An investigation of helicity injection by photospheric shear motions is carried
out for two active regions(ARs), NOAA 11158 and 11166, using line-of-sight magnetic
field observations obtained from the Helioseismic and Magnetic Imager on-board
Solar Dynamics Observatory. We derived the horizontal velocities in the active
regions from the Differential Affine Velocity Estimator(DAVE) technique.
Persistent strong shear motions at the maximum velocities in the range of
0.6--0.9km/s along the magnetic polarity inversion line and outward flows from
the peripheral regions of the sunspots were observed in the two active regions.
The helicities injected in NOAA 11158 and 11166 during their six days'
evolution period were estimated as $14.16\times10^{42}$Mx$^2$ and
$9.5\times10^{42}$Mx$^2$, respectively. The estimated injection rates decreased
up to 13\% by increasing the time interval between the magnetograms from 12 min
to 36 min, and increased up to 9\% by decreasing the DAVE window size from $21\times18$
to $9\times6$ pixel$^2$, resulting in 10\% variation in the accumulated helicity. 
In both ARs, the flare prone regions (R2) had inhomogeneous helicity flux distribution with 
mixed helicities of both signs and that of CME prone regions had almost homogeneous 
distribution of helicity flux dominated by single sign. The temporal profiles of 
helicity injection showed impulsive variations during some flares/CMEs 
due to negative helicity injection into the dominant region of positive 
helicity flux. A quantitative analysis reveals a marginally significant 
association of helicity flux with CMEs but not flares in AR 11158, while for 
the AR 11166, we found marginally significant association of helicity flux with 
flares but not CMEs, providing evidences of the role of helicity injection at 
localized sites of the events. These short-term variations of helicity flux are
further discussed in view of possible flare-related effects. This study suggests
that flux motions and spatial distribution of helicity injection are important
to understand the complex nature of magnetic flux system of the active region
leading to conditions favorable for eruptive events.


\end{abstract}

\keywords{Sun: activity --- Sun: flares --- Sun: magnetic fields---
Sun:Coronal Mass ejections--- Sun: helicity injection}

\section{Introduction}
\label{Intro}

Magnetic helicity is an important topological property of solar active regions (ARs) and
is a measure of twist and writhe of the field lines \citep{berger1984, finn1985}.
It is gauge invariant for a closed volume of space. The Sun's outer
atmosphere is dominated by magnetic field at all scales. Dynamic phenomena, such as,
energetic flares and coronal mass ejections (CMEs) occur due to the loss of equilibrium
during the evolution of magnetic fields in solar ARs. Magnetic helicity has become an
important physical parameter in the context of solar transient phenomena. It is one of
the few global quantities which is conserved even in resistive magneto-hydrodynamics on a
timescale less than that of the global diffusion. There exists no absolute measure of
helicity within a sub-volume of space if that sub-volume is not bounded by a magnetic
surface. However, a topologically meaningful and gauge invariant relative helicity for
such volumes can be defined.

There are several methods for estimating helicity in solar ARs. By
the force-free field assumption of coronal magnetic field, we have:

\begin{equation}
    \nabla \cdot \mathbf{B}=\alpha \mathbf{B}\label{eq_lff}
\end{equation}

\noindent where $\alpha$ is the force-free parameter, also known as helicity or twist
parameter. Assuming $\alpha$ to be constant for the whole AR, we can fit observed vector
magnetograms to deduce the value of $\alpha$ \citep{pevtsov1995,hagyard1999,sanjiv2009}.
Latitudinal variation of helicity of photospheric magnetic fields, and statistical
significance of the observed temporal variations of the ARs' hemispheric helicity rule,
as measured by the latitudinal gradient of the best-fit linear force-free parameter
$\alpha$, etc., have been discussed by \citet{pevtsov2008}.

The Poynting-like theorem for helicity in an open volume as derived by \citet{berger1984}
is given by:

\begin{equation}
\frac{dH}{dt}=\oint 2(\mathbf{B_{t}}\cdot\mathbf{A_p})v_z ds - \oint 2(\mathbf{A_p}\cdot
\mathbf{v})B_{z}ds \label{eq_heli_ber}
\end{equation}

\noindent where $\mathbf{A_p}$ is the vector potential of the potential
magnetic field, $\mathbf{B_p}$, which is uniquely specified by the observed
flux distribution on the surface(x-y plane) as

\begin{equation}
\nabla \times \mathbf{A_p} \cdot \hat{z}=B_{z};\hspace{.3cm} \nabla \cdot
\mathbf{A_{p}}=0;\hspace{.3cm} \mathbf{A_{p}}\cdot \hat{z}=0 \label{eq_bc}
\end{equation}

\noindent where $\hat{z}$ refers to unit vector along vertical direction of
Cartesian-geometry. Equation~\ref{eq_heli_ber} shows that the helicity of
magnetic fields in an open volume may change by the passage of helical field
lines through the surface (first term) and/or by photospheric footpoint motions
of the field lines (second term). The temporal evolution of magnetic helicity
flux across the photosphere characterizes the injection of magnetic helicity
from the sub-photospheric layers into the solar atmosphere, horizontal flux
motions, and the changes in the coronal magnetic field configurations related to
eruptive events, such as the CMEs, propagating into the interplanetary medium.

During the past years, several attempts have been made to estimate magnetic helicity from
suitable solar observations. \cite{chae2001a} developed a method for determining the
helicity flux (the second term in Equation~\ref{eq_heli_ber}) passing through the
photosphere. They used a time series of photospheric line-of-sight (LOS) magnetograms to
determine horizontal velocities by local correlation tracking (LCT) technique
\citep{november1988}. Using this method, vector potential $\mathbf{A_p}$ was constructed
by using photospheric LOS field (as an approximation to $B_z$ field) as boundary conditions
with Coulomb gauge in terms of Fourier Transform (FT) as:
\begin{eqnarray*}
A_{\rm p,x}&=&FT^{-1}\left[\frac{jk_y}{k_{x}^{2}+k_{y}^{2}}FT\left(B_{\rm z}\right)\right]  \\
A_{\rm p,y}&=&FT^{-1}\left[\frac{-jk_x}{k_{x}^{2}+k_{y}^{2}}FT\left(B_{\rm z}\right)\right]
\label{eq_fou_sol}
\end{eqnarray*}

\noindent where $k_x$, $k_y$ are spatial frequencies in the x, y directions,
respectively. Later, this method was applied to many ARs by several authors
\citep{chae2001b,moon2002,nindos2003,chae2004}. However, \citet{pariat2005}
showed that this method of calculation introduced artificial polarities of both
signs in the helicity flux density maps with many flow patterns.
Therefore, they suggested to use relative velocities for calculating the
helicity injection rate:

\begin{equation}
\frac{dH}{dt}=\frac{-1}{2\pi}\int_{S}\int_{S'}\frac{[(\mathbf{x}-\mathbf{x'})\times(\mathbf{u}-\mathbf{u'})]_n}{|\mathbf{x}-\mathbf{x'}|^2}B'_z(\mathbf{x'})
B_z(\mathbf{x}) dS'dS \label{eq_par}
\end{equation}

\noindent where $\mathbf{u}$ is the foot-point velocity at the position vector
$\mathbf{x}$, and $B_z$ is the vertical component of the observed magnetic field.
This equation shows that the helicity injection rate can be understood as the
summation of relative rotation rates of all the pairs of elementary fluxes weighted with
their magnetic flux.

Furthermore, \citet{schuck2005} has shown that the LCT method is inconsistent with the
magnetic induction equation, which governs the temporal evolution of the photospheric
magnetic fields. Tracking methods have serious practical limitations that might result
in the failure of detecting significant shear velocity fields and hence in the underestimate
of the amount of helicity injected by such velocity fields. \citet{demoulin2003} reported
that the magnetic energy and helicity fluxes should be computed only from the horizontal
motions deduced by tracking the photospheric cross-section of magnetic flux tubes.
These authors contend that the apparent horizontal motions include the effect of both the
emergence and the shearing motions. They analyzed the observational difficulties involved in
deriving such fluxes and in particular, the limitations of the correlation tracking methods.
One of the main limitations in the previous studies has been the coarse spatial resolution
of the available observations which limits the deduced velocities to the velocity corresponding
to the group motion of an unresolved bunch of thin flux tubes covered by a pixel. Also,
tracking motions faced difficulties in the areas lacking sufficient contrast,
such as in the sunspot umbrae.

Several alternative, improved methods have been developed  for inferring plasma
velocities consistent with the induction equation at the photospheric level, based on the
LOS, as well as, vector magnetograms. The Induction method (IM; \citealt{kusano2002}),
induction local correlation tracking(ILCT; \citealt{welsh2004}), minimum energy fit (MEF;
\citealt{longcope2004}), differential affine velocity estimator (DAVE;
\citealt{schuck2005,schuck2006}) and differential affine velocity estimator for vector
magnetograms (DAVE4VM; \citealt{schuck2008}) have been developed for the determination of
horizontal component of motion. In contrast, the normal component of velocity can
be determined from the doppler observations of ARs located near the disk center. DAVE4VM
method requires vector magnetograms. The performance of different
techniques has been examined in \citet{welsh2007} which showed that the MEF, DAVE,
FLCT, IM, and ILCT algorithms performed comparably. Furthermore, they reported that while
the DAVE estimated the magnitude and direction of velocities slightly more accurately
than the other methods, MEF's estimates of the fluxes of magnetic energy and helicity
were more accurate.

Time series data of photospheric magnetograms have been extensively used to derive
magnetic helicity and its evolution in order to examine its role in the level of
transient activity of the ARs. \cite{moon2002} reported impulsive variations of helicity
during some M and X-class flares. In a survey, \citet{labonte2007} revealed that
X-flaring ARs have a higher net helicity change with peak helicity rate
$>6\times10^{36}$Mx$^2$s$^{-1}$ with weak hemispheric preference.
\citet{park2010a} have also studied the solar flare productivity in relation to the
helicity injection using a large sample of 378 active regions. Using SOHO-MDI
magnetograms, they reported variation of helicity injection rates and a significant
helicity accumulation of $(3-45 \times 10^{42})$ Mx$^2$ over several days around the time
of flares above M5.0. Most of the previous studies that used data from Michelson  Doppler
Imager (MDI) onboard SOHO had the time resolution of 96 minutes. The rather coarse time
resolution between two consecutive observations has been a matter of concern in the above
calculations because the contribution from the motion of short lived magnetic features in
small intervals is difficult to be accounted suitably ({\it e.g.}, \citealt{chae2004}). This
underlines the need for observations of magnetic fields with higher temporal resolution.

The above mentioned issues can now be addressed with the availability of a better cadence
of 12 minutes by the recently launched Helioseismic and Magnetic Imager (HMI) onboard
{\it Solar Dynamics Observatory} (SDO). Our main objective in the present work is to
reinvestigate the role of helicity injection in relation with flares and CMEs using the
high-resolution data obtained from SDO-HMI. We intend to utilize this opportunity to
revisit some of the previous studies involving computations of helicity rate for two ARs,
NOAA 11158 and NOAA 11166, that appeared during February and March 2011, respectively, in
the ascending phase of the current Solar Cycle 24.

Using the high quality HMI data obtained for the two ARs, we intend to examine whether
the variations as reported by \cite{moon2002} and \citet{park2010b} for energetic flares
occurred also during the flares of lower magnitude. It is of particular interest to
investigate if such changes were associated with the CMEs as well. For our analysis,
we use DAVE technique for retrieving horizontal foot-point velocity from the LOS magnetograms.
Thereafter, using Equation~\ref{eq_par} we determine helicity injection rates and
the accumulated helicity in the two ARs due to foot-point shearing motions during their
disk transit. It has been inferred from the previous studies that most of the helicity
injection corresponds to magnetic flux emergence in the ARs\citep{jeong2007}. We, therefore,
attempt to interpret the variations found in these physical parameters in relation to the
occurrence of flares and CMEs. In particular, we investigate whether the transport of magnetic
helicity plays a role in solar eruptions.

We organize this paper as follows. Description of the data  used in
this study and the procedures of the data processing are given in
Section~\ref{Data}. Results obtained for the two selected ARs are
presented in Section~\ref{Results} and the following discussions in
Section~\ref{disc}. The summary of the work presented in this article
is given in Section~\ref{Sum}.

\section{Data and Method of Analysis}
\label{Data} For our study, we have used high resolution LOS magnetograms at a cadence of
12 minutes obtained from the Helioseismic Magnetic Imager (HMI; \citealt{schou2012})
on board {\it Solar Dynamic Observatory} (SDO). HMI observes the full solar disk in the Fe
{\sc i} 6173\AA~ spectral line with a spatial resolution of 1 arc-second. HMI provides
four main types of data: dopplergrams (maps of solar surface LOS velocity), continuum
filtergrams (broad-wavelength maps of the solar photosphere), LOS and vector magnetograms
(maps of the photospheric magnetic field).

NOAA 11158 (19$^\circ$S) and 11166 (10$^\circ$N) appeared on the
disk during February 11-20, 2011 and March 03-16 2011 respectively.
These ARs were very active, and produced some intense X-class flares
associated with CMEs in addition to many M- and C- class flares
during their disk transits. From the AIA observations, intermittent
mass expulsions were seen, many of which turned into large, fast
moving CMEs as further confirmed by STEREO. Table~\ref{TabFlCm} gives a
list of flares\footnote{Obtained from the website
\url{www.solarmonitor.org}} (as recorded by GOES) and
CMEs\footnote{by scrutinizing AIA 304\AA~ quicklook images mirrored
at \url{http://jsoc.stanford.edu/data/aia/images/2011/} and further
confirmed by the timings from
\url{http://spaceweather.gmu.edu/seeds/}}.

\begin{table*}
 \centering \caption{List of Flares and CMEs}
\begin{tabular}{l l l l}
  \hline                                                                                      \\
  AR     &  Date      &  Flares   &   CMEs                                               \\
  (NOAA)   &dd/mm/yyyy        & magnitude(time UT)& (time UT)                                           \\
  \hline                                                                                   \\
  11158    & 11/02/2011 & No flares & No CMEs                                               \\
           & 12/02/2011 & No flares & No CMEs                                                \\
           & 13/02/2011 & C1.1(12:29),C4.7(13:44),M6.6(17:28) & 21:30,23:30                   \\
           & 14/02/2011 & C1.6(02:35),C8.3(04:29),C7.1(06:51) & 02:40,07:00,12:50,17:30,19:20        \\
           &            & C1.8(08:39),C1.7(11:51),C9.4(12:41) &                                 \\
           &            & C7.0(13:47),M2.2(17:20)*,C6.6(19:23) &                                       \\
           &            & C1.2(23:14), C2.7(23:40)             &                                     \\
           & 15/02/2011 & X2.2(01:44),C2.7(00:31)             & 00:40,02:00*,03:00,04:30,05:00      \\
           &            & C4.8(04:27),C4.8(14:32),C1.7(18:07) & 09:00                         \\
           &            & C6.6(19:30),C1.3(22:49)             &                               \\
           & 16/02/2011 & C2.0(00:58),C2.2(01:56),C5.9(05:40) & 14:35                    \\
           &            & C2.2(06:18),C9.9(09:02),C3.2(10:25) &                                 \\
           &            & C1.0(11:58),M1.0(01:32),M1.1(07:35) &                                        \\
           &            & M1.6(14:19),C7.7(15:27),C1.3(19:29) &                                         \\
           &            & C1.1(20:11),C4.2(21:06),C2.8(23:02) &                                  \\
                                                                                                \\
  11166    & 06/03/2011 & C5.1(11:56),C3.9(15:21)             & 02:00,15:20                       \\
           & 07/03/2011 & M1.9(13:45)                         & 14:25,22:10                     \\
           & 08/03/2011 & C7.7(23:10)                         & 14:30,19:00                    \\
           & 09/03/2011 & C9.4(08:23),M1.7(10:35),M1.7(13:17) & 06:40,10:40,21:45      \\
           &            & C9.4*(22:03),X1.5(23:13)              &                                          \\
           & 10/03/2011 & C2.9(03:50),C6.2(07:03),C4.2(13:19) &                                   \\
           &            & C4.7(13:42),C2.0(14:21)              & 04:50,07:10                    \\
           &            & C4.0(19:00),M1.1(22:34)              &                                  \\
           & 11/03/2011 & C1.4(00:29),C1.1(01:46),C2.8(02:24)  & 00:50                          \\
           &            & C5.5(04:15),C4.3(07:22),C1.1(08:13)  &                                \\
           &            & C2.0(11:10),C3.6(11:43),C1.1(16:04)  &                                  \\
           &            & C1.0(22:20),C1.0(22:50)              &                         \\
\hline                                                                                            \\
\end{tabular}                                                                                  \\
{\it Note:} All flares(CMEs) associated to source region R2(R1) of respective ARs except those
marked by *\label{TabFlCm}
\end{table*}

Magnetograms obtained at different times were aligned by the method of
\citet{chae2004}. In this method, an image of the AR taken at the central
meridian is considered as the reference image. All other images,  in time
accounted for differential rotation \citep{howard1990} along with the
latitudinal difference of center of reference image from the ephemeris
information, were remapped on to the disk center. This method is intended to
reduce errors due to geometrical foreshortening and the AR is transformed to
the disk center. Since at disk center, normal and vertical 
components of magnetic fields are same, the difference between the normal and LOS component 
was corrected by cosine of the distance of the AR center from the disk center by
assuming the horizontal field contribution for the transformation to be
negligible \citep{venkat1989}.

We followed the transits of the two selected ARs from east to west on the solar disk. In
order to have negligible errors in geometric correction, we restricted ourselves to a
region within $\pm$40$^\circ$ longitude from the central meridian. With this
constraint, we confined our study of the temporal evolution of the ARs to six days's
period around their central meridian passage.

We derived the horizontal velocities of foot-points on
the photosphere by using DAVE technique \citep{schuck2006}.  The DAVE technique
is essentially a local optical flow method that determines the magnetic
footpoint velocities within the windowed region. Further, it adopts an affine
velocity profile specifying velocity field in the windowed region about a point
and constrains that profile to satisfy the induction equation. Any tracking
method depends on two parameters, {\it viz.}, the window size and the time
interval. For a given time interval $\Delta$t, the window should be large
enough so that tracked features remain confined within the window. Also, it
should be small enough to be consistent with an affine velocity profile.
\citet{schuck2008} presented a way to select an optimal window objectively,
using the degree of consistency between change in the observed magnetic field
($\Delta{\rm B}/\Delta t$) and the expected magnetic field change based on
the flow estimated with several trial windows. They found the best performance
of this method at approximately a square window of pixels. Since the ARs
evolved rapidly, we chose a window size of 21$\times$18 pixel$^2$ after a
careful verification of the physical flux motions and directions of estimated
flows. The dependence of helicity injection rate on window size and time
difference between the tracked maps using this method were investigated. Moreover,
as the HMI magnetic field measurement precision is 10G \citep{schou2012}, we
have set this as the threshold to avoid errors while retrieving velocities.
Further details of this method are given in a recent work of \citet{tian2011}.

Computation of the helicity rate using the method (direct integration) proposed
by \citet{pariat2005} at each pixel of the AR map ({\it cf.}, Equation
\ref{eq_par}) is a tedious, time consuming process. However, we chose to use
this method for reducing the effect of fake polarities of helicity flux. Restricting 
the calculations at pixels with magnetic field above the threshold
($\geq$10G) helps to reduce the computation time typically by 15-25\%.
Parallelization in integrand computation further reduces the time approximately
by a factor of the number of processors used. The same equation as rewritten by
\cite{chae2007} to suit the convolution algorithm by Fourier transform is
faster than the direct integration method. The intrinsic problem of Fourier
transform with periodicity could be overcome by padding the array corresponding
to the data points with rows and columns of zeros to get results as obtained by
direct integration method. In this study, we have implemented the former
approach (direct integration) to get sufficiently accurate results.

\section{Evolution of Magnetic Flux and Helicity}

\label{Results} The evolution of observed magnetic flux and the computed helicity  rates
are presented in the following for the two selected ARs NOAA 11158 and NOAA 11166 with
the methods and procedures explained in Section~\ref{Data}.

\subsection{AR NOAA 11158}
\label{a158}

\begin{figure*}
\centering
\includegraphics[width=.99\textwidth,clip=,bb=60 37 524 270]{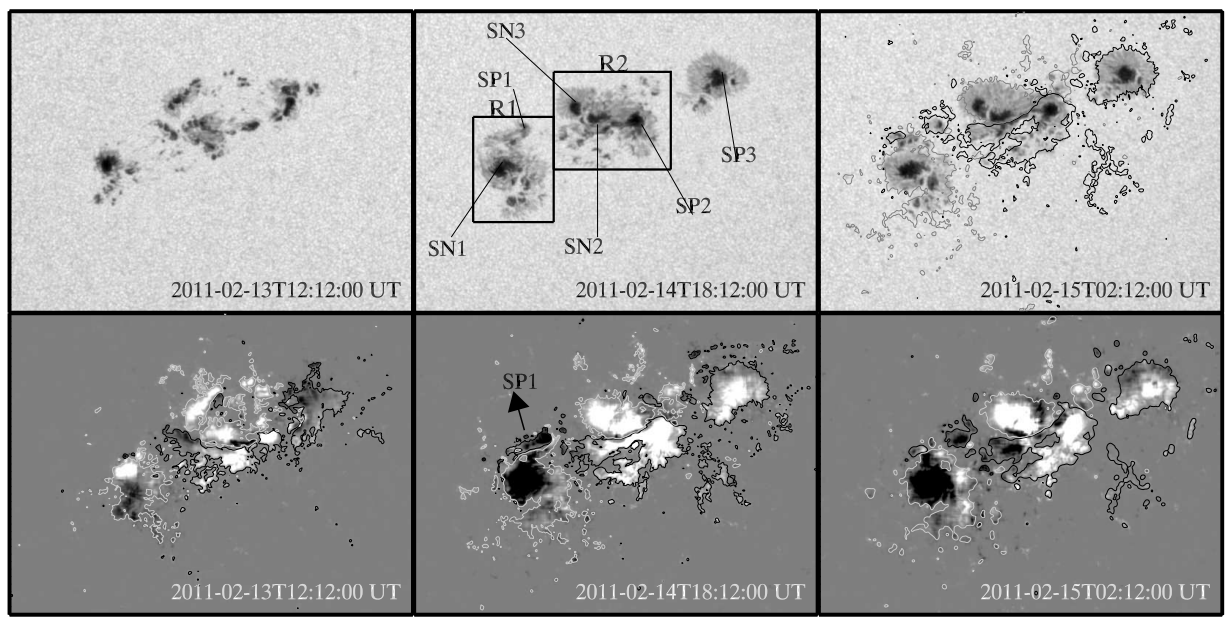}
\caption{(\emph{Top row}) The daily HMI continuum intensity maps of AR  NOAA
11158, and (\emph{Bottom row}) the corresponding helicity flux density maps
(scaled to $\pm0.05\times10^{20}$Mx$^2$cm$^{-2}$s$^{-1}$ and also in subsequent
plots) computed from Equation~\ref{eq_par}. The field of view is $275 \times
200$ arcsec$^2$. The overlaid gray and black contours correspond to LOS
magnetic fields at [-150,150]G levels, respectively. Rectangular boxes in
intensity image of 2011 February 14 mark the selected sub-areas R1 and R2 in
which velocity flows are shown in the subsequent figures.} \label{58_mos}
\end{figure*}

This AR appeared as small pores at the heliographic location E33S19 on 2011
February 11 as seen in the full disk HMI photoheliograms. Thereafter, it grew
very rapidly during the next two days as the small pores merged and formed
bigger sunspots. It was a newly emerging region which developed to a large AR
having $\beta\gamma\delta$ magnetic complexity during its rapid evolution. It
consisted of four large regions of opposite polarities in quadrupolar
configuration. Figure~\ref{58_mos}(top row) shows the evolution of NOAA 11158
during 2011 February 13-15 in HMI intensity maps. The prominent positive
polarity sunspots of the AR are labeled as SP1, SP2, SP3 and the negative polarity spots
as SN1, SN2, SN3 for identification. LOS contours are overlaid on the intensity
image showing the respective polarity distribution.

The spatial evolution of the AR shows a large shearing motion of SP2 that rotated around
SN2 about its umbral axis during 2011 February 13-15. It then detached and moved towards
SP3 along with small patches of both polarities appearing and disappearing over short
periods of time. This motion appeared to have created a twist in magnetic fieldlines
connecting these spots. A careful examination of the animation made from magnetograms and
intensity maps revealed a significant counter clock-wise (CCW) rotation of SN1 during the
same period, while a small positive polarity region SP1 located to the north of SN1
rotated in the counter clock-wise direction along with a proper motion away from SN1. The
rotations of SN1 and SP1 increased the twist of the field lines, and the magnetic
non-potentiality of the sigmoid structure \citep{canfield1999}. Several mass expulsions
were launched intermittently from this region, as seen from the quick look images in AIA.
These turned into CMEs as confirmed by STEREO observations.

In order to quantitatively analyze the magnetic complexity or twist contributed
by the observed shearing motions of the magnetic foot points, we computed the
helicity injection rates using the temporal sequence of magnetograms of the AR.
Figure~\ref{58_mos}(bottom row) shows the computed helicity flux density maps
corresponding to the HMI continuum intensity images (top row). The dark
(white) patches in the right panel represent negative (positive) helicity flux
density according to the usual convention. Contours of LOS magnetic field at
[-150, 150]G levels are overlaid for a better visualization of helicity flux
density with respect to the magnetic polarity. Evidently, negative polarity
region of SN1 injected negative (dark) helicity during 2011 February 14-15
which is also consistent with its physical CCW rotation. In contrast,
SP2, SN2 and SN3 injected positive (white) helicity along with negative (dark)
helicity in some small patches. We expect that the nature of motions in these
areas could have influenced the helicity pattern there.

The photospheric maps of helicity flux (and its injection rate) provides spatial
information about the basic properties of a link between the activity and its
sub-photospheric roots as reflected by the flux emergence process. In a sample of four 
active regions, \citet{jeong2007} found that helicity was mostly injected while fluxes 
emerged in the AR, suggesting it to be the major source of helicity injection. The flux 
cancellation process, on the other hand, resulted in a loss of coronal magnetic helicity, or 
inverse helicity injection. We thus infer that the AR possessed two main sites, of unstable 
energy storage systems marked by the rectangular boxes R1 and R2 in Figure~\ref{58_mos}. These 
sites had distinctly different injection of helicity flux density corresponding to the flux (or
foot-point) motions, polarities and activity.

\begin{figure*}
\centering
\includegraphics[width=0.99\textwidth,clip=,bb=2 1 488 505]{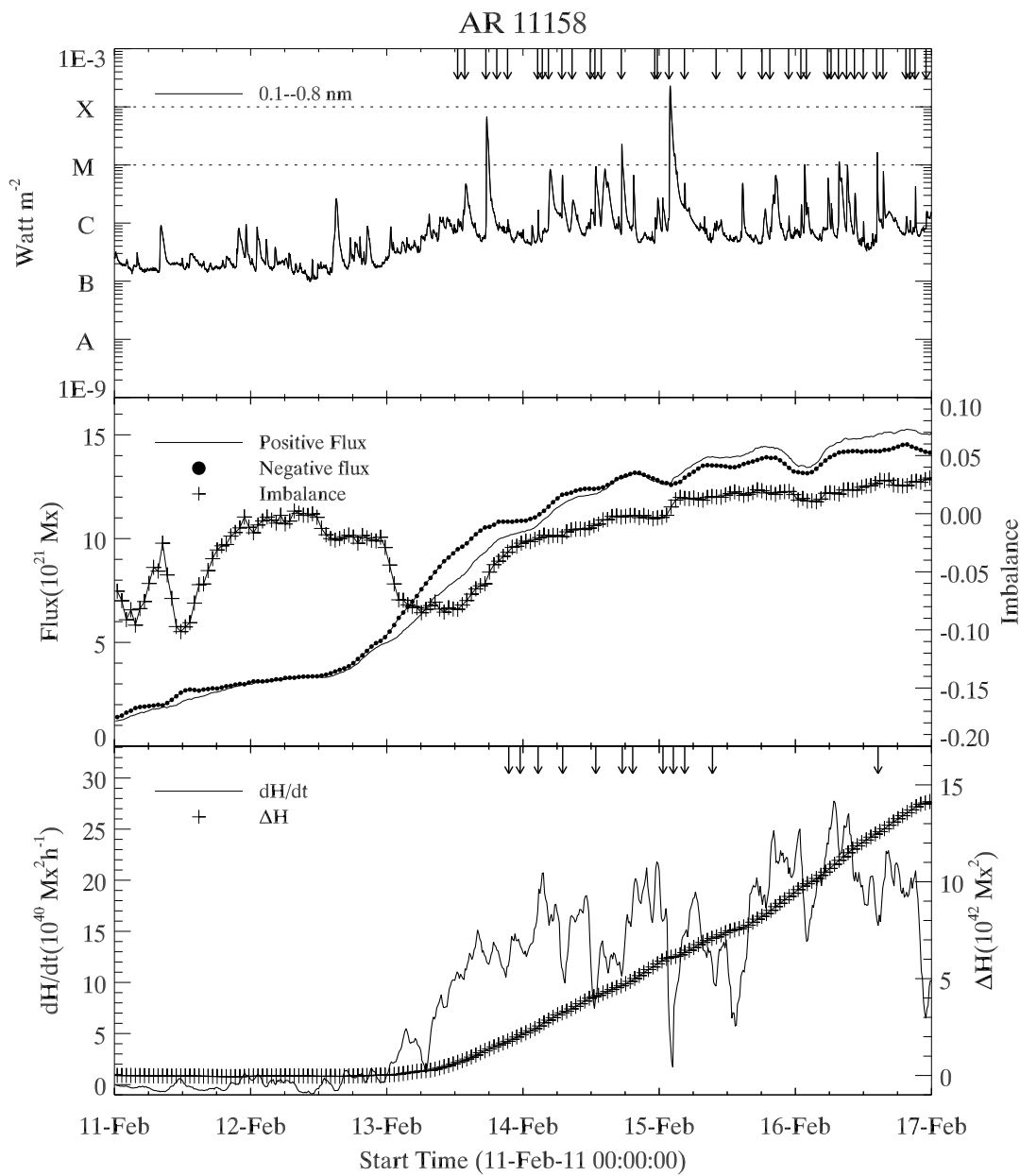}
\caption{\emph{Top}: Solar disk integrated GOES Soft X-ray flux during February
11-16, 2011. The arrows on top panel indicate the start times of flares in AR
NOAA 11158. \emph{Middle}: Time profiles of the magnetic fluxes and flux
imbalance in the AR. \emph{Bottom}: The computed helicity rates integrated over
the whole AR. Arrows in this panel indicate the onset time of CMEs that were
launched from this AR.}\label{58_tplot}
\end{figure*}

In order to show the transient activity of the AR as it evolved, we have
plotted the disk integrated GOES soft X-ray flux (1-8\AA~channel) during
February 11--17 in Figure~\ref{58_tplot}(top) where the start times of flares
of NOAA 11158 are marked by arrows. After its birth, the AR gradually evolved
during 2011 February 11--13 as evident from the monotonic increase of fluxes in
both polarities corresponding to $3\times10^{21}$ Mx (Figure~\ref{58_tplot},
middle). Then followed a rapid phase of flux emergence (of $9\times10^{21}$ Mx)
during February 13--14 after which it reached a plateau. Also plotted is the
flux imbalance, {\it i.e}, the ratio of the net flux and absolute total flux in
the AR. The dominance of negative flux during February 13--15, and thereafter
of the positive flux, is evident. Flux variations occurred in the range of
(9.5--12.5)$\times 10^{21}$ Mx with the imbalance within $10\%$ over six days.
A significant flux decrease in both polarities by $\sim1\times10^{21}$Mx
occurred till the time of the X2.2 flare. We shall discuss more about flux
changes during X-flares in Section~\ref{disc}. The unusual rotating sunspots
along with the increased fluxes indicated emergence of highly twisted fluxes
from the sub-photospheric region \citep{leka1996}, and not resulting from the
surface flows alone. Most of the flare and CME activity of this AR occurred
only after February 13/12:00UT, indicating that the rapid flux emergence could
have played important role in triggering the transients.

In Figure~\ref{58_tplot}(bottom), we have plotted the time profile of helicity
injection rate($dH/dt$), which is the summation of helicity flux density over
the AR . Also plotted is the accumulated helicity, {\it i.e}, the integrated
helicity change rate over time ($\Delta H=\int\frac{dH}{dt}\Delta t$). The
total accumulated helicity is estimated as 14.16$\times10^{42}$Mx$^2$ during
the six day period of 2011 February 11--16, with the peak helicity rate of
31.54$\times10^{40}$Mx$^{2}$h$^{-1}$. The occurrence times of the CMEs
associated with the AR are marked by arrows in this panel for reference. An
impulsive variation of helicity injection rate due to injection of negative
helicity is discernible during the X2.2 flare and the concomitant CME. The
helicity injection rate decreased during the period February 14/11:00--February
15/13:00 UT, and increased thereafter till February 16 along with fluctuations
in the range 2--4$\times10^{40}$Mx$^{2}$h$^{-1}$. We notice a large dip of
helicity injection around X2.2 flare with associated CMEs. We have smoothed the
original time profile at 12 minute interval by a box car window of five data
points ({\it i.e.}, 1 hour). Similar sudden dips in injection rates during
other events can be further analyzed for examining their association.

\begin{figure*}
\centering
\includegraphics[width=.99\textwidth,clip=,bb=61 38 538 205]{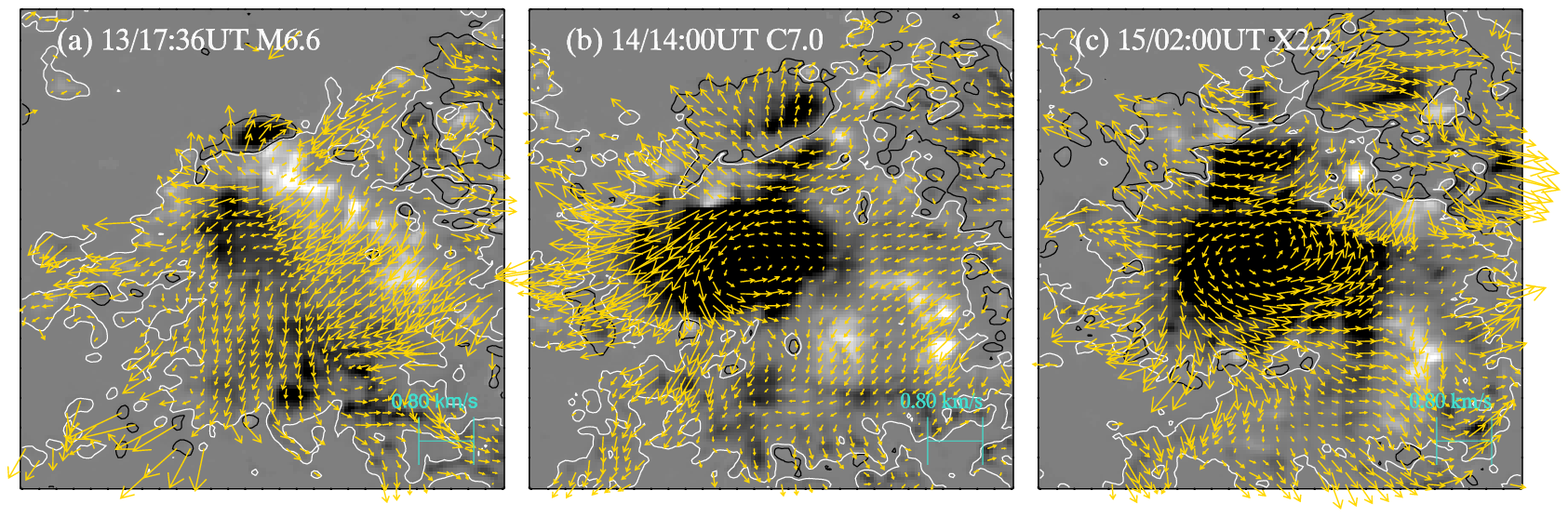}
\includegraphics[width=.99\textwidth,clip=,bb=64 37 534 163]{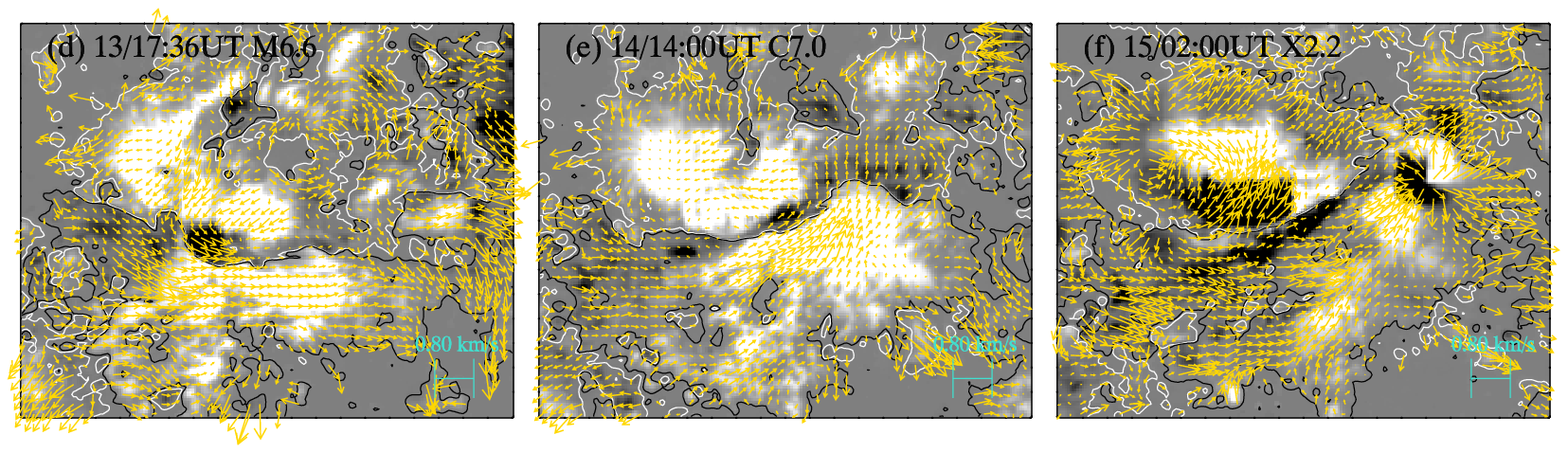}
\caption{Transverse velocity field vectors as inferred from DAVE technique
superposed on helicity flux density maps with the LOS magnetic field contours
for the rectangular regions of Figure~\ref{58_mos} -- R1 (\emph{Top row}) and
R2 (\emph{Bottom row}). Spiral or vortex like velocity patterns in sunspot penumbra 
in (b-c) are due to umbral rotation of sunspot SN1. Sites of 
negative helicity injection are seen around the magnetic polarity inversion line
in (d)-(f) at the peak times of the flares noted in each panel.}
\label{58_vel}
\end{figure*}

Figure~\ref{58_vel} shows transverse velocities in the  rectangular sub-regions
R1 (top row) and R2 (bottom row) of NOAA 11158 overlaid on the corresponding
maps of helicity flux density during three flare events. Also overlaid are the
contours of the LOS magnetic flux at $\pm150$G levels. Maximum rms velocities
in the range of 0.6--0.9\kmps~were found over the observed period in the AR.
Spiral or vortex like velocity patterns are obviously related to the counter
rotation of SN1 in Figure~\ref{58_vel}(b--c). A notable observation is that the
sub-region R1 possessed negative helicity flux density distribution which is
consistent with the chirality associated with the physically observed counter
rotation of SN1 whereas R2 possessed mixed helicity flux dominated by positive
helicity flux distribution. Because of the continued shearing motions at the
interface of SP2 and SN2, the flow field vectors almost aligned with the
polarity inversion line (PIL) as seen in panels (d--f). Interaction of fluxes
with this shear motion can squeeze and converge the flux in both SP2 and SN2.
We hypothesize that the field lines were stressed and twisted by this motion
leading to the storage of free energy adequate to account for the release in
the energetic X2.2 flare of February 15/01:44UT.  As almost all flares (except
M2.2 at 14/17:20) occurred in R2, we examined the spatial distribution of
helicity flux before and after the flare events to know whether any sudden
changes are found related to the occurrence of flare. During some events, we
noticed negative patch of helicity flux in the regions of positive helicity
flux. Especially, in the panels (e--g), a negative helicity flux distribution
near the PIL during M6.6, C7.0, and X2.2 flares can be observed. There may be
some concern about these flare-related changes, as it is known that during the
impulsive phase of large flares, the spectral line profile itself may undergo
some change affecting the magnetic (and velocity) field measurements.

\begin{figure*}
\centering
\includegraphics[width=.99\textwidth,clip=,bb=45 16 595 396]{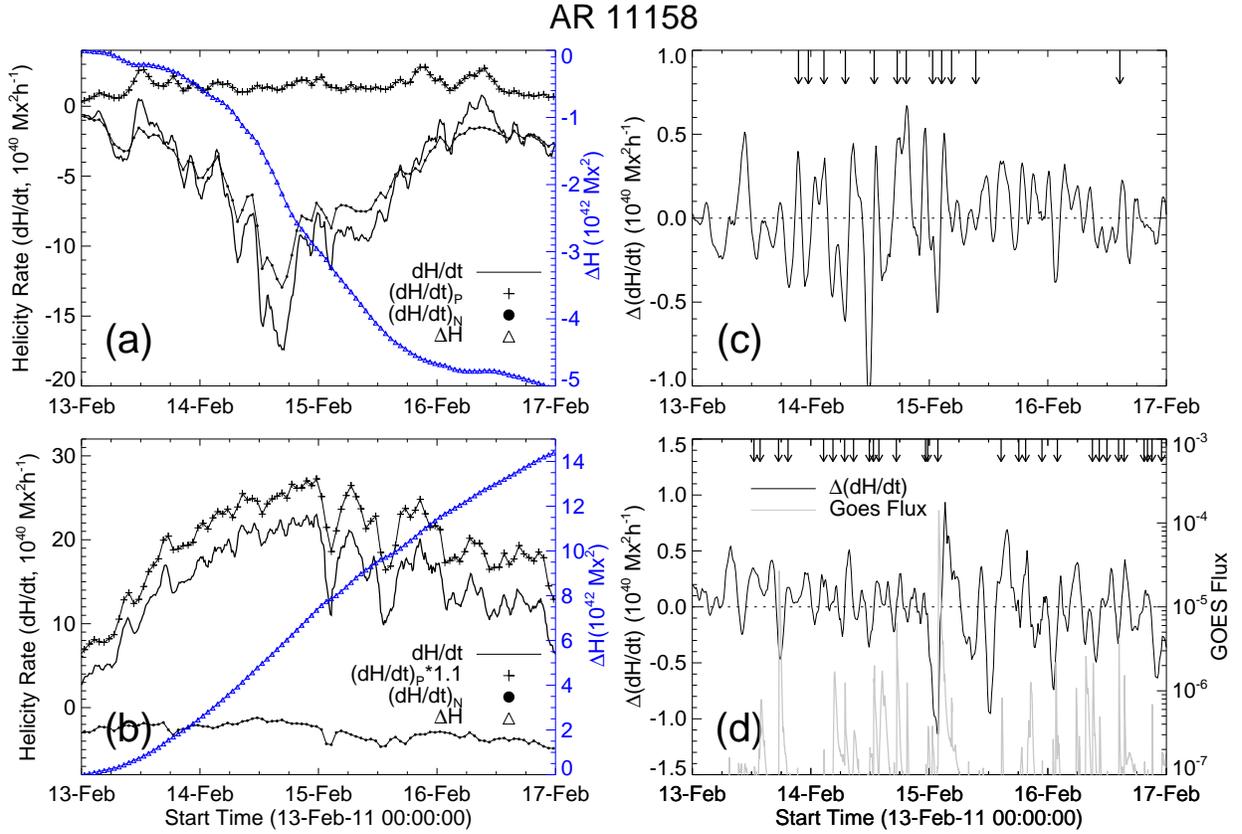}
\caption{Temporal evolution of helicity rate and accumulated helicity
integrated over (a) R1 and (b) R2. The time difference of helicity
rate($\Delta(dH/dt)$) in (c) for region R1 with arrows marking CME timings,
(d) for region R2 with pointed flares originated from this AR.}\label{roi_58}
\end{figure*}

Most of these flares occurred in R2 while the mass expulsions(or CMEs) were
associated to R1. In order to relate helicity rate changes to these events,
therefore, we have computed and plotted the total injected quantities for R1
and R2 in Figure~\ref{roi_58}(a-b). Injection of helicity in a region of
dominant opposite sign can be understood as a sudden dip in the time profile
plot. Of course, the corresponding spatial information is lost in the averaged
quantity. The advantage of using localized analysis of selected sub-areas in
the ARs is that it reduces complex variations occurring over a much larger area
of the entire AR while showing only the variations occurring in the
areas-of-interest. It also enhances the dips corresponding to the identified
events (marked by the arrows). However, it is important to identify the
location of individual event in order to correctly attribute a particular
change of helicity rate to it. NOAA 11158 was essentially a positive helicity
injecting region, while its sub-region R1 had a negative injection rate and
accumulated quantity due to the presence of rotational motion. We expect that
as the sunspots SN1 and SP1 rotated, the injection rate increased to a maximum
of $-16\times10^{40}$Mx$^2$h$^{-1}$ on February 14/18:00UT. A total helicity
accumulation of $-5.60\times10^{42}$Mx$^{2}$ occurred during the six day period
in this region. Noticeably, a steep accumulation occurred during Feb 14--15
along with many observed mass expulsions shown by arrows. This could be
interpreted as shedding of excess helicity from the corona in the form of
eruptive events. The steep accumulation of helicity by monotonic injection
rate, therefore, is suggested to be a cause of expulsions. Accumulated
helicity amounting to $14.44\times10^{42}$Mx$^{2}$ in sub-region R2 with steep
accumulation observed from February 13 onwards, could be mostly associated with
the observed large shear motion of SP2.

For a quantitative study of the association of short term variations in
helicity rate to the flaring or CME, the following analysis is carried out.
The absolute time difference of the helicity flux ($|\Delta(dH/dt)|$, having units same 
as dH/dt) averaged over start and stop times of GOES flares above C2.0 is computed.
This is compared to that of randomly selected but equal length time intervals
containing no flares. A significantly higher mean of $|\Delta(dH/dt)|$ during
flares compared to quiet times would indicate a robust association between
flaring and helicity fluxes. A similar analysis is undertaken for time windows
around CMEs to look for a CME-helicity flux association. We assume
that there is no time lag between flaring and helicity flux signal while
carrying out this analysis. We first interpolated the signal at 1 min interval
from 12 min interval to get values as required by the GOES flare times, then
it was smoothed to a boxcar width of 30 minutes. Within start and stop times of flares,
the averaged value of absolute variation was computed to compare with that
calculated during randomly selected, constant interval(30 min) quiet times.

The time difference of helicity rate in R1-R2 is shown in
Figure~\ref{roi_58}(c-d) with CMEs and flares marked by arrows. Large amplitude
variations are discernible during M6.6, X2.2 and the CME at 12:30UT indicating
some association, but similar variations are present around the mean position
even in quiet times. From the above described analysis, we found a significantly 
higher mean during CME's ($0.054\pm0.007$) compared to quiet times($0.032\pm0.008$).
The difference in CME versus quiet time helicity fluctuations are marginally 
statistically significant, at better than one-sigma. Similarly, a mean of $0.044\pm0.004$($0.049\pm0.008$) 
during flare (quiet) times indicate poor or no association of flaring to 
helicity flux variations. The same analysis for the helicity flux over the 
entire AR improved the association (in terms of mean absolute helicity variation) 
slightly for CMEs but worsened it for flaring. We shall further discuss 
these helicity variations during flare/CMEs in view of the involved flare-related 
effects in Section~\ref{FlrEff}.

\subsection{AR NOAA 11166}
\label{a166}

\begin{figure*}
\centering
\includegraphics[width=.99\textwidth,clip=,bb=60 37 524 270]{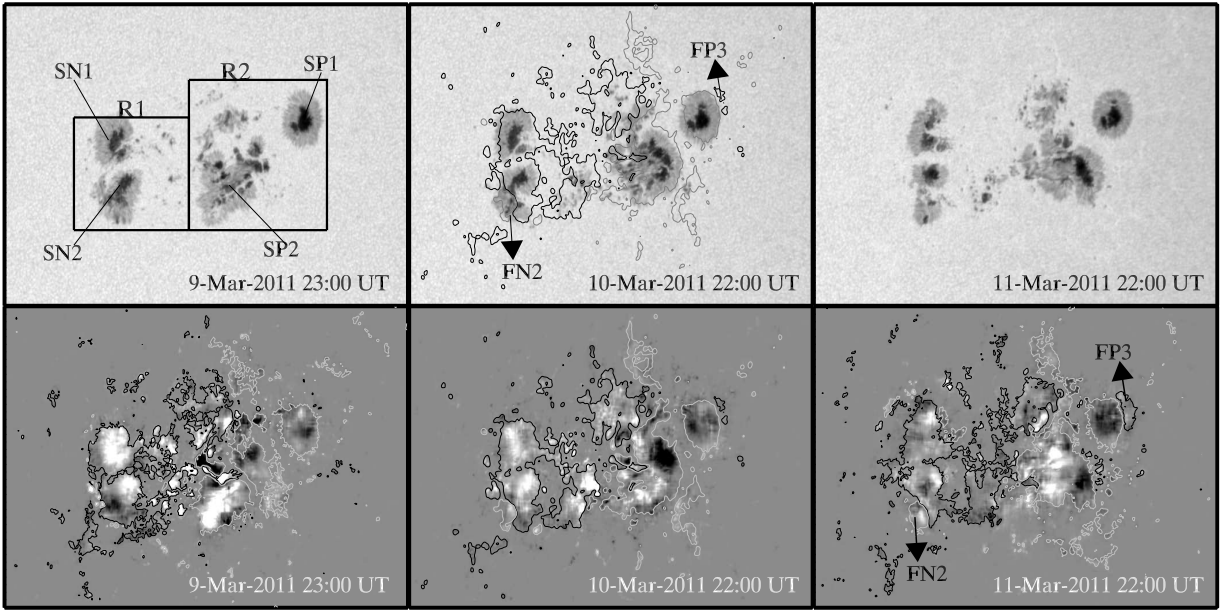}
\caption{(\emph{Top row}) The daily HMI continuum intensity maps of AR NOAA
11166, and (\emph{Bottom row}) the corresponding helicity flux density maps
computed from Equation~\ref{eq_par}. The field of view is $350 \times 200$
arcsec$^2$. The overlaid gray and black contours correspond to LOS magnetic
fields at [-150,150]G levels, respectively. Rectangular boxes in intensity
image of March 9 mark the selected sub-areas in which velocity flows are shown
in the next figure. Emerging fluxes from sunspot periphery are indicated as FN2
and FP3 on March 11/22:00UT }\label{66_mos}
\end{figure*}

AR NOAA 11166 appeared on the east limb of solar disk on 2011 March 03 at the
location N10E64. We monitored its activity during the period of 2011 March
6--11 in which it produced a large X1.5 flare, two M-class flares and several
C-class flares, some of which were also associated with plasmoid ejections or
CMEs. Table~\ref{TabFlCm} lists the flares and CMEs of this AR. Daily evolution
of the AR in the period of March 8-11, 2011 is shown in Figure~\ref{66_mos}(top row).

The major sunspots of the AR are labeled as SP1, SP2, SN1 and SN2. The
identification of SP2 was somewhat unclear before March 10 as several
small umbrae were spread over its location. They moved and coalesced to
form SP2. Polarities of the respective sunspots are identified by the
overlaid LOS magnetic field contours. This AR also consisted of a
complex magnetic configuration with two positive (SP1, SP2) and two
negative (SN1, SN2) polarity sunspots located within the surrounding
diffused fluxes. Emerging and moving flux regions, FP3 and FN2, were
identified in the course of the evolution in the sunspot periphery
(March 11/22:00UT panel), having opposite sign to that of their native
sunspots. However, there were no intrinsic rotating sunspots or flux
patches as observed in the case of AR NOAA 11158.

We computed the helicity flux density for AR NOAA 11166 during its evolution
in the period 2011 March 6--11. The corresponding maps for three successive
days are plotted in Figure~\ref{66_mos}(bottom row). Locations of helicity
flux density of mixed sign were distributed all over the AR through out
the evolution period. The  peripheral sites of the sunspots exhibited
helicity flux density predominantly of negative sign. However, patches
of negative helicity flux were also observed embedded in the positive
helicity flux site of the flare (March 09/23:00UT panel). For further close
examination, we consider two sub-areas R1 and R2, as marked by the boxes
in this panel.

\begin{figure*}
\centering
\includegraphics[width=1.0\textwidth,clip=,bb=2 1 490 505]{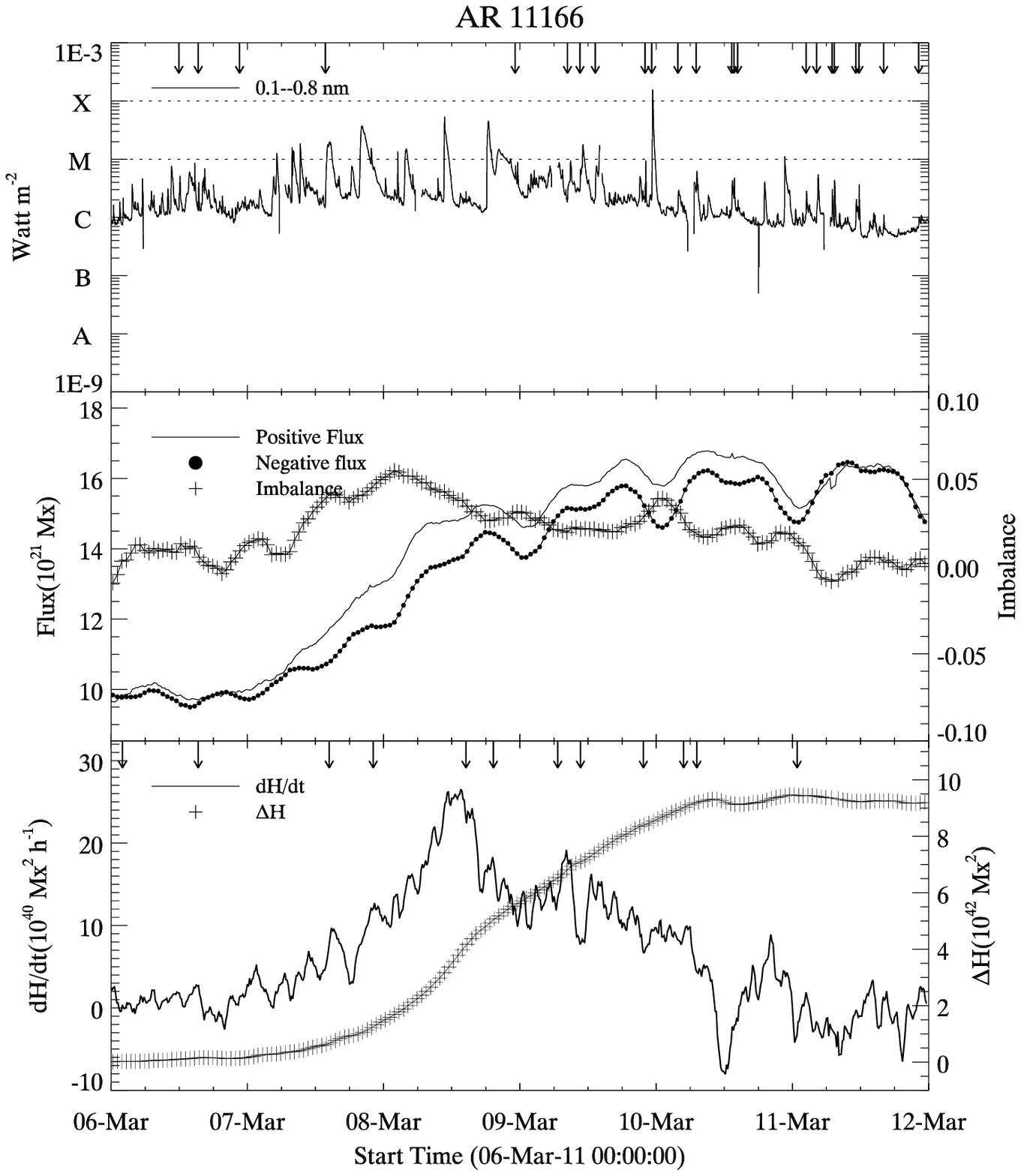}
\caption{Same as Figure~\ref{58_tplot} but for AR NOAA 11166.} \label{66_tplot}
\end{figure*}

The disk integrated GOES soft X-ray flux (1-8\AA~channel) during 2011 March 6-11
is plotted in Figure~\ref{66_tplot}(top). The arrows in this panel indicate the start
time of flares in NOAA 11166. During the disk transit of the AR, fluxes of both
polarities increased corresponding to $5\times10^{21}$Mx, with the imbalance varying
below 6\% (Figure~\ref{66_tplot}, middle). As observed for NOAA 11158, a rapid flux emergence
occurred in this AR too during March 7--9. Thereafter, only small variations associated
with local cancellations/emergence of about $\sim1\times10^{21}$Mx took place pertaining
to the gradual evolution of the AR. Positive flux dominated in the AR during March 7-11,
and then a near balance was established. It is worth noticing that magnetic fluxes
in both polarities decreased by $\sim0.9\times10^{21}$Mx while evolution of fluxes
leading to the occurrence of a CME following the X1.5 flare.  However, it is not clear
whether this decrease in flux six hours before the flare/CME has some role in these events.
But, the flux imbalance, increasing prior to the flare, reduced significantly after the flare 
consistent with observations reported by \citet{wang2010}. Most of the flares and CME activity 
of this AR occurred only after March 8, suggesting that the rapid emergence of fluxes could be 
an important factor for triggering of these transients.

Temporal evolution of helicity injection rate and the accumulated helicity for
NOAA 11166 are shown in Figure~\ref{66_tplot}(bottom) with arrows marking the
times of the  CMEs. A five magnetogram boxcar was used to smooth the profile to reduce
fluctuations in the profile. As expected, these parameters increased in the first phase
corresponding to the flux emergence, in agreement with \citet{jeong2007} that
helicity is mostly injected while the fluxes emerged. Total helicity accumulated
during the six days' period of the AR's evolution was estimated to $\sim9.5\times10^{42}$Mx$^{2}$.
The maximum helicity injection occurred during 2011 March 8 at the rate of
$30\times10^{40}$Mx$^2$h$^{-1}$. Thereafter, it reduced gradually to the
minimum rate at $-10\times10^{40}$Mx$^2h^{-1}$ on 2011 March 10.
The coronal helicity of the AR is likely to be positive as a result of this
positive helicity injection.

\begin{figure*}
\centering
\includegraphics[width=0.99\textwidth,clip=,bb=64 37 505 162]{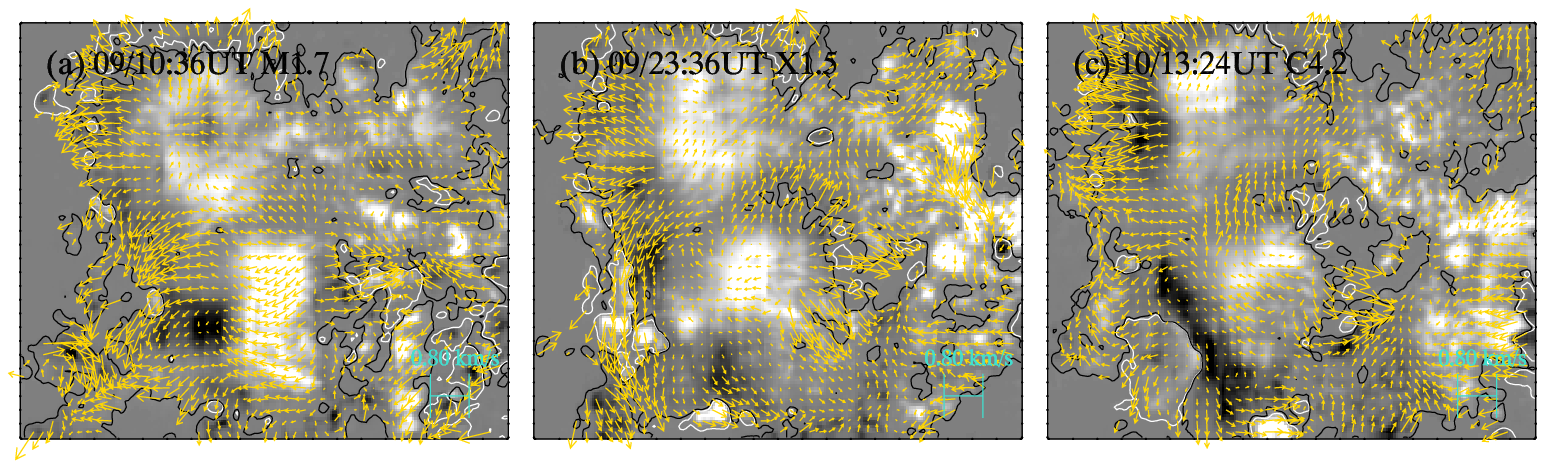}
\includegraphics[width=0.99\textwidth,clip=,bb=64 37 505 162]{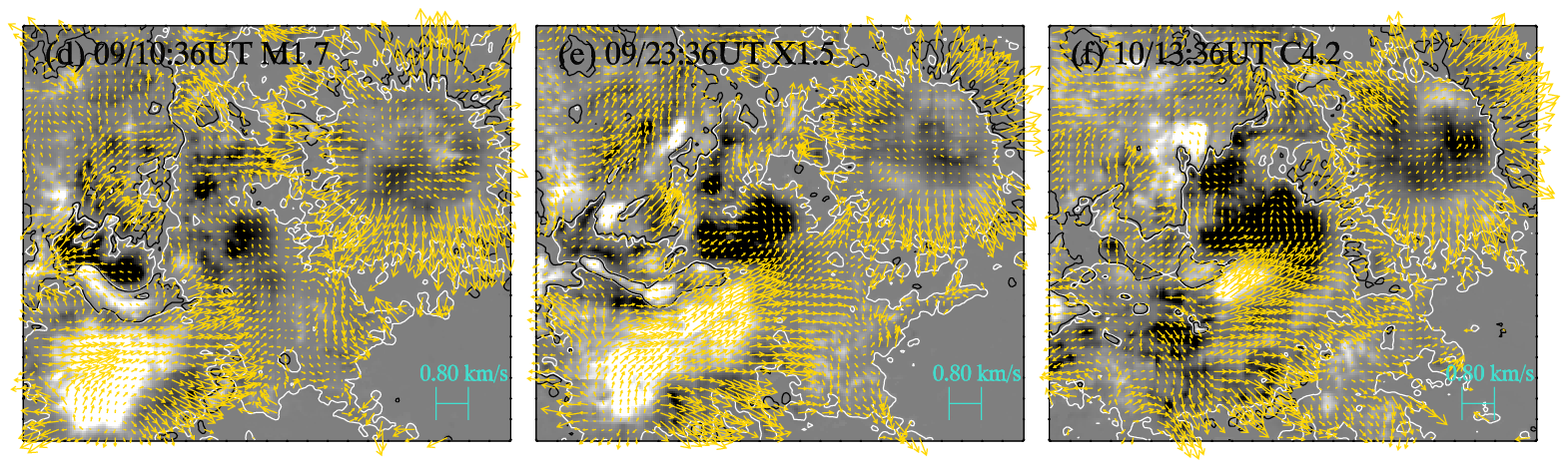}
\caption{Transverse velocity field vectors in the rectangular region R1
(\emph{Top row}) and R2 (\emph{Bottom row}) of Figure~\ref{66_mos}
overlaid on the helicity flux density maps with iso-contours of LOS
magnetic field during flare events.} \label{66_vel}
\end{figure*}

Horizontal, or transverse, velocity vectors corresponding to the tracked flux
motions are plotted in Figure~\ref{66_vel} separately for R1 (top row) and
R2(bottom row). The rms velocities of flux
motions are found to have the maximum values in the range 0.5--0.9~\kmps.
Strong moat flows were systematically dominant in both regions from the
peripheral regions of sunspots in addition to the shear flows. Persistent
strong shear motions due to the merging SP2 group were identified in R2.
These flows appear to collide head on with those from SP1 resulting in the flux
submergence/cancellation. Flux emergence was also identified from the diverging flow field
observed in animated flows from R1. From this region, flux moved towards R2 as
the AR evolved. Such motions appear to be associated with injection of negative
helicity into a region with predominantly positive flux, increasing the complexity of
the magnetic flux system as shown in panels (d)--(f) of R2. Further, these negative
helicity injections often coincided with some observed events, such as the
three of them shown in this plot. For the X1.5 flare the distribution of
helicity flux is shown in panel (e) on March 09/23:36UT.

\begin{figure*}
\centering
\includegraphics[width=.99\textwidth,clip=,bb=45 16 595 396]{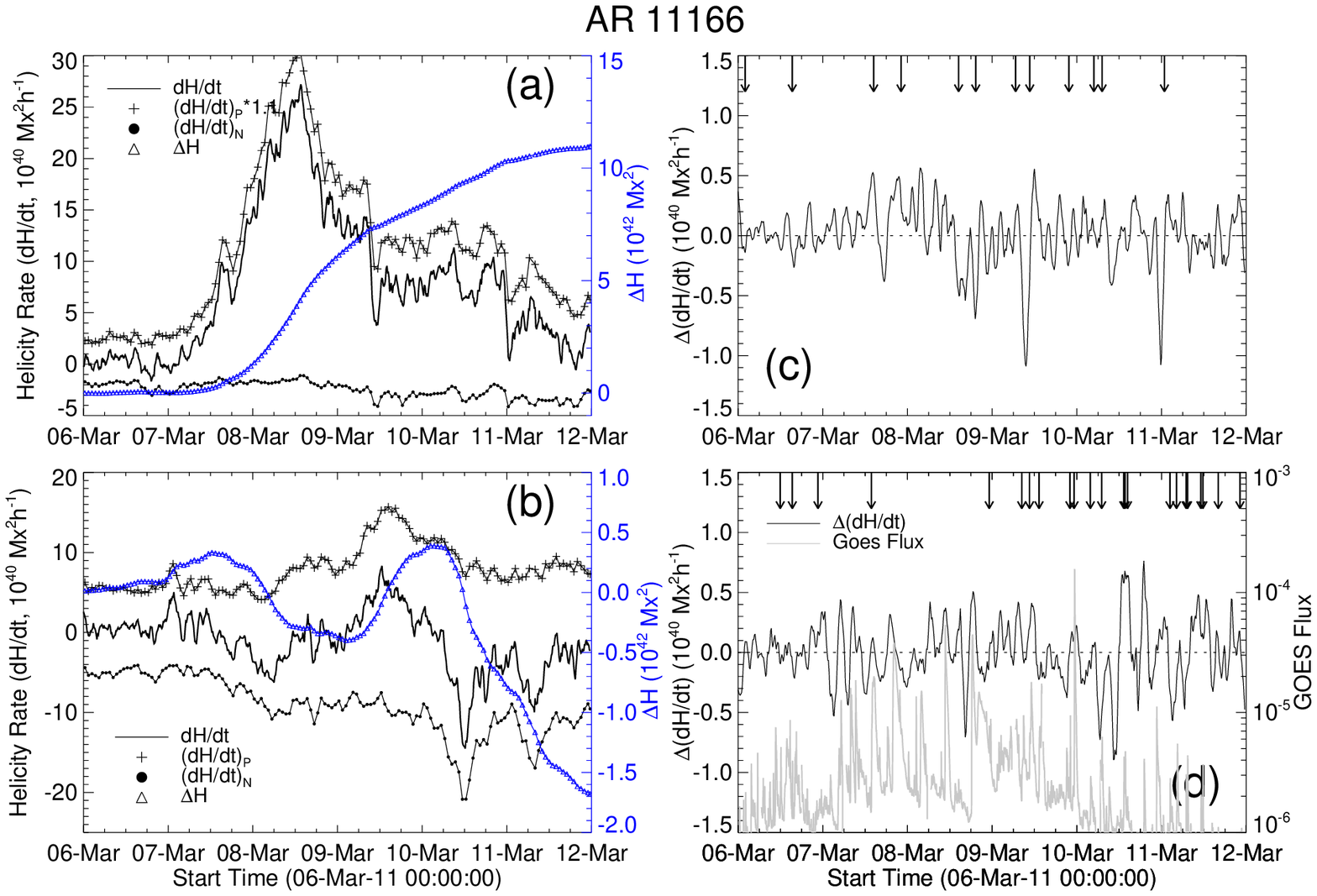}
\caption{Same as Figure~\ref{roi_58} but for AR NOAA 11166.} \label{roi_66}
\end{figure*}

The injection rates and accumulated helicities deduced from sub-regions R1 and
R2 are plotted in Figure~\ref{roi_66}(a--b). Also the contribution of each
signed helicity flux in the net helicity flux is plotted separately. The time
profile of R1 shows it to have positive helicity injection with a steep
increasing phase during March 7--9 at a peak rate of 
$27\times10^{40}$Mx$^2$h$^{-1}$. Thereafter, gradual decrease in the rate of
injection is evident from the plot. As mentioned earlier, R1 was a site of
emerging flux that resulted in contributing to accumulation of helicity
amounting to $11\times10^{42}$Mx$^2$. While R2 exhibited mixed
sign injection rates during its evolution. As in the previous AR, continuous
injection of dominant positive helicity from R1 is suggested to be the cause of
observed mass expulsions, whereas the injection from R2 is of mixed signs
suggested to result in flares. An enhanced peak of helicity rate was seen
around the time of the X1.5 flare in R2 of AR 11166 that was not obvious in
Figure~\ref{66_tplot}(bottom panel) since we reduced fluctuations
occurring over entire AR by selecting small area. After this event, the negative 
injection rate increasingly dominated on March 10, turning the net injection of the entire 
AR negative. The implication of this transition of injection rate
from positive to negative sign over a day is not clear in the observed events
shown by the arrows.

The time variation of helicity flux in both R1 and R2 are plotted in
Figure~\ref{roi_66}(c--d) along with the arrows pointing start times of CMEs
and flares in the AR. Some of the large amplitude variations of helicity flux
about the mean position appear to be related to these events. As in the
previous AR, we have analyzed the association of flare/CMEs that originated
from the sub-regions R1 and R2 of this AR with the respective helicity flux.
The calculated mean of variation in helicity flux ($|\Delta (dH/dt)|$) during 
flaring ($0.099\pm0.020$) is marginally statistically different at about two-sigma level 
over that during quiet times ($0.057\pm0.007$), reflecting a robust association of 
flaring and helicity fluxes. The mean of $|\Delta(dH/dt)|$ obtained in quiet 
times do not have any information or bias of flaring or CME, therefore higher mean during the
flare/CMEs implies some impact of helicity flux variations in them.
A similar analysis undertaken for CMEs also showed the similar association(
during CMEs of $0.052\pm0.006$ dominated over quiet times of
$0.047\pm0.006$, but not statistically significant difference). However, the 
association strengthened for flaring and weakened for CMEs when the helicity flux over the 
entire AR was considered in the analysis.

\subsection{Flare-related effects on Helicity flux}
\label{FlrEff} It is well known that the photospheric magnetic (and Doppler)
field measurements are affected by flares. During an energetic flare, the
profile of spectral line used for the measurement was reported to change from
absorption to emission, resulting in a change of sign in the deduced magnetic
polarity \citep[and references therein]{qiu2003}. This abnormal polarity
reversal was observed to last for about a few minutes during the impulsive
phase of the flare (typically 3-4 minutes).  Similar abnormal, transient
changes have also been reported for some other large, white light flares
\citep{maurya2009,maurya2012}. The change in the line profile may arise due to
both thermal effects and non-thermal excitation and ionization by the
penetrating electron jets produced during the large flares. We term these as
flare-related transient changes, considered to be artifacts as they do not
correspond to real magnetic field changes.

There is increasing evidence that flares may change the magnetic field more
significantly on a persistent and permanent manner
\citep{sudol2005,petrie2010,wang2010}. The persistence of the observed field
changes implies that they are not artifacts of changes in the photospheric
plasma parameters during the flare, and the temporal and spatial coincidences
between flare emission and the field changes suggest the link of the field
changes to the flare. We term these as permanent flare-related changes. With
these known transient and permanent flare-related effects on magnetic fields,
it would  not be clear,  particularly during the impulsive phase of the flare,
if the change in helicity flux can be interpreted as genuine transport of
helicity across the photosphere.

In addition, an implicit assumption made in our approach of calculating
helicity injection is the ideal evolution of photospheric magnetic fields in
the induction equation used to derive velocities of flux motions. Moreover, the
same assumption is involved in the derivation of helicity injection from the
relative helicity formula \citep{berger1984,finn1985}. This assumption is valid
and reasonable outside the flaring time intervals (at least during permanent
changes of fields) as the typical observed photospheric velocities are far less
than the Alfven velocities. In the real conditions of rapid, transient changes
in photospheric magnetic fields spanning impulsive period of the flare,  the
assumption of ideal magnetic evolution may not be applicable. Therefore, there
is theoretical uncertainty regarding the interpretation of helicity fluxes
during flares.

\begin{figure*}
\centering
\includegraphics[width=.98\textwidth,clip=,bb=61 37 559 414]{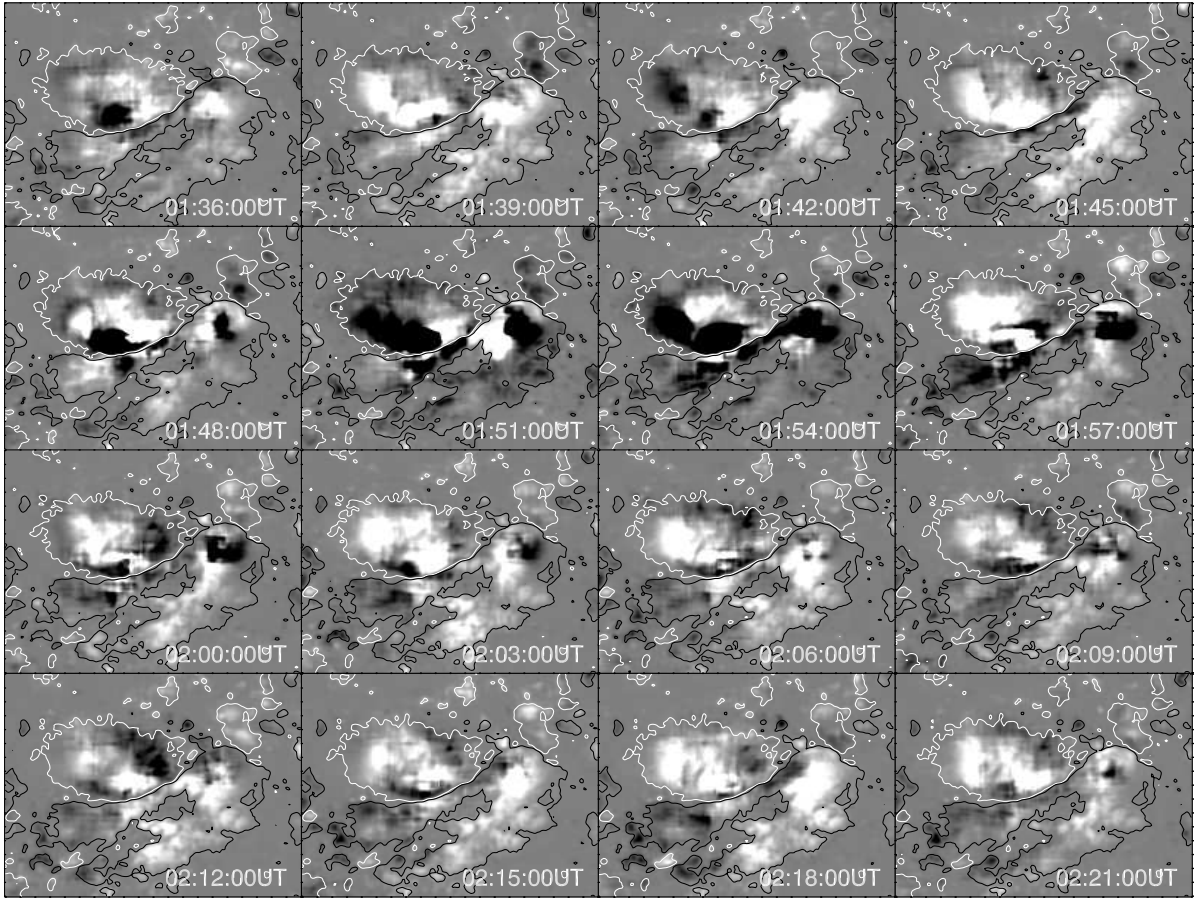}
\caption{Mosaic of injection of helicity flux distribution around the time of
X2.2 flare in AR 11158 with iso-contour of LOS positive(negative) flux in black(white).
Intense negative helicity flux about the PIL during peak time(01:48--02:00UT)
of the flare is evident possibly due to flare-related transient effect on the magnetic
field measurements during the impulsive period.}\label{MosX22}
\end{figure*}

In order to inspect these aspects in the signal of the helicity change rate, we
procured 45s cadence magnetograms for some selected flare events and averaged
them to 3min cadence after processing as the previous data set. A mosaic of
distribution of helicity flux around the X2.2 flare is shown in
Figure~\ref{MosX22}. During the impulsive period (01:48-02:02UT) of this flare,
negative helicity flux is distributed about the PIL which we believe to be due
to the transient flare-related effect. The magnetic (and Doppler) transients
and locations of spectral line reversal associated with this flare are already
reported by \citet{maurya2012}, which are spatially and temporally consistent
with this negative helicity flux distribution. Therefore, the observed negative
helicity flux distribution in the dominant positive site can be attributed to
the transient flare-effect,  and is likely to be artifact, i.e., not a true
transfer of helicity.

\begin{figure*}
\centering
\includegraphics[width=.325\textwidth,clip=,bb=66 365 368 708]{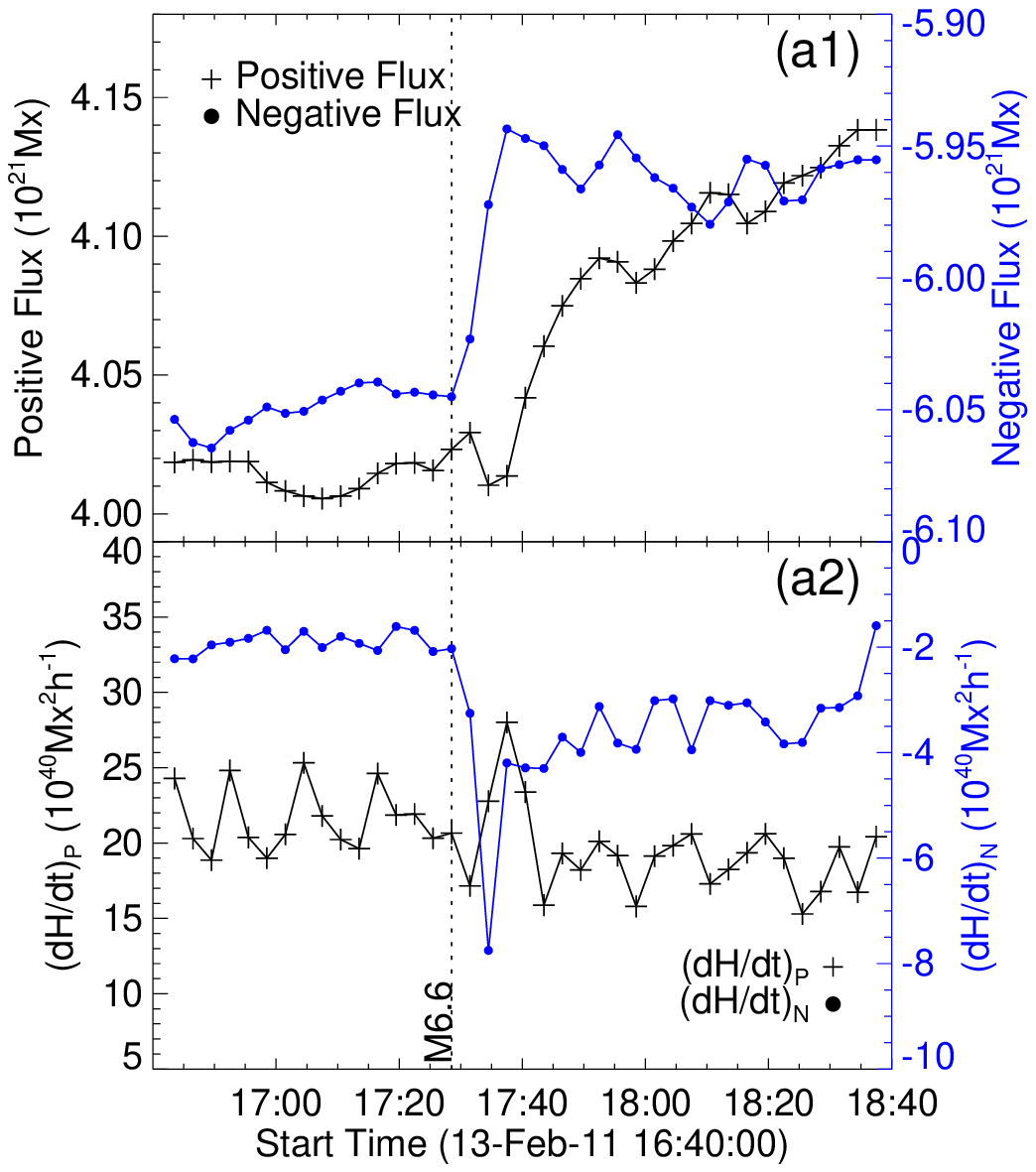}
\includegraphics[width=.325\textwidth,clip=,bb=66 365 368 708]{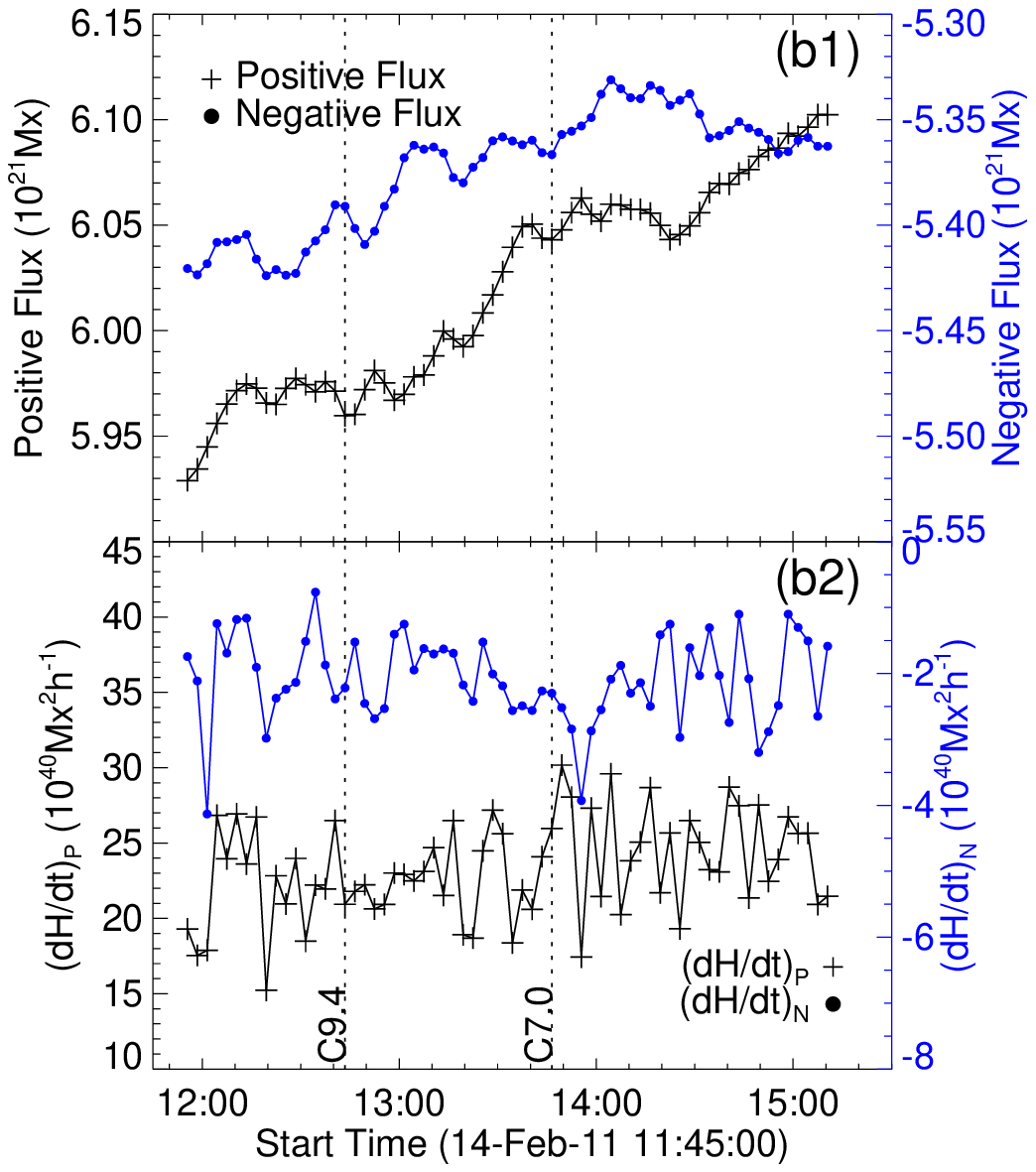}
\includegraphics[width=.325\textwidth,clip=,bb=66 365 368 708]{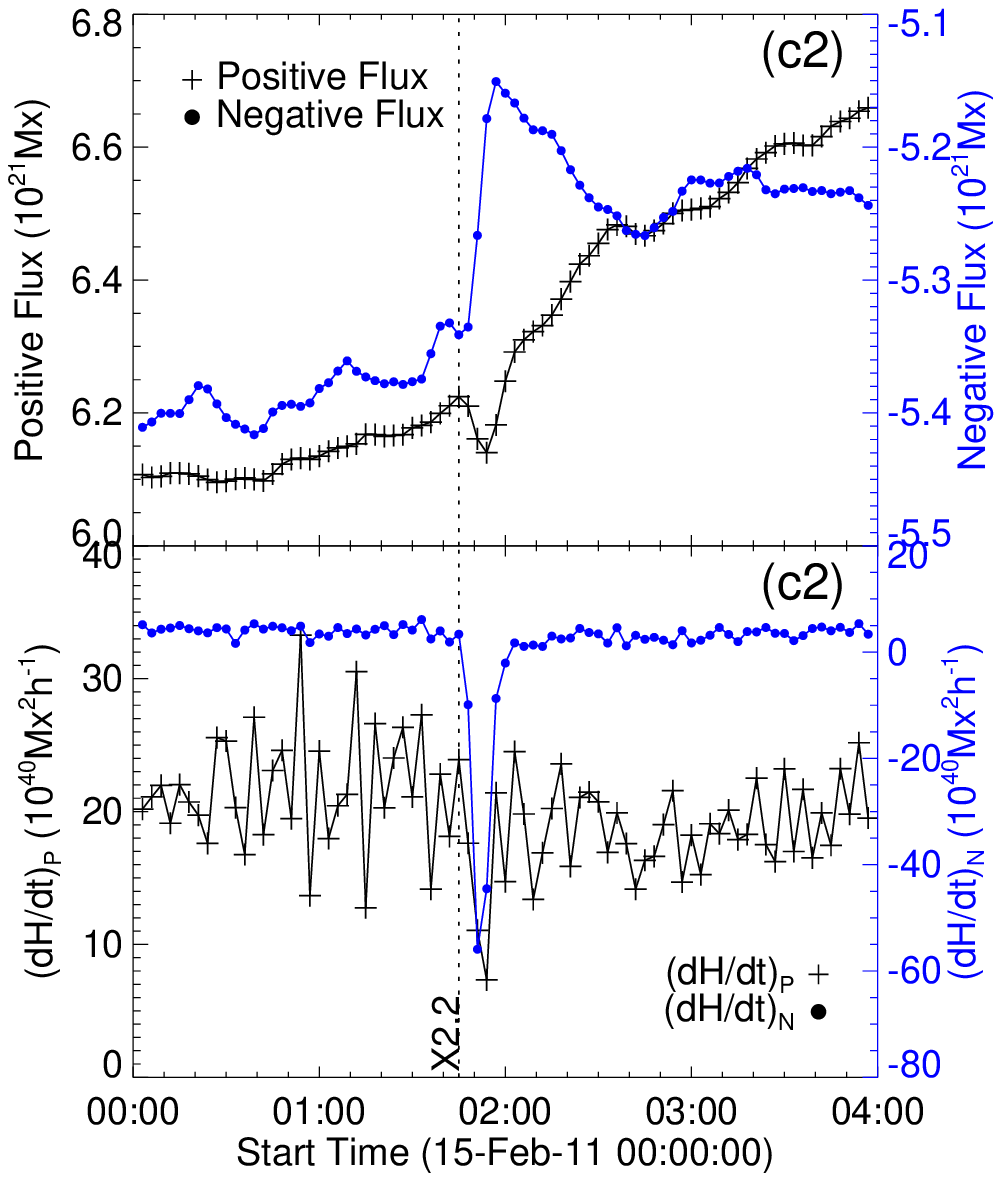}
\includegraphics[width=.325\textwidth,clip=,bb=66 365 368 708]{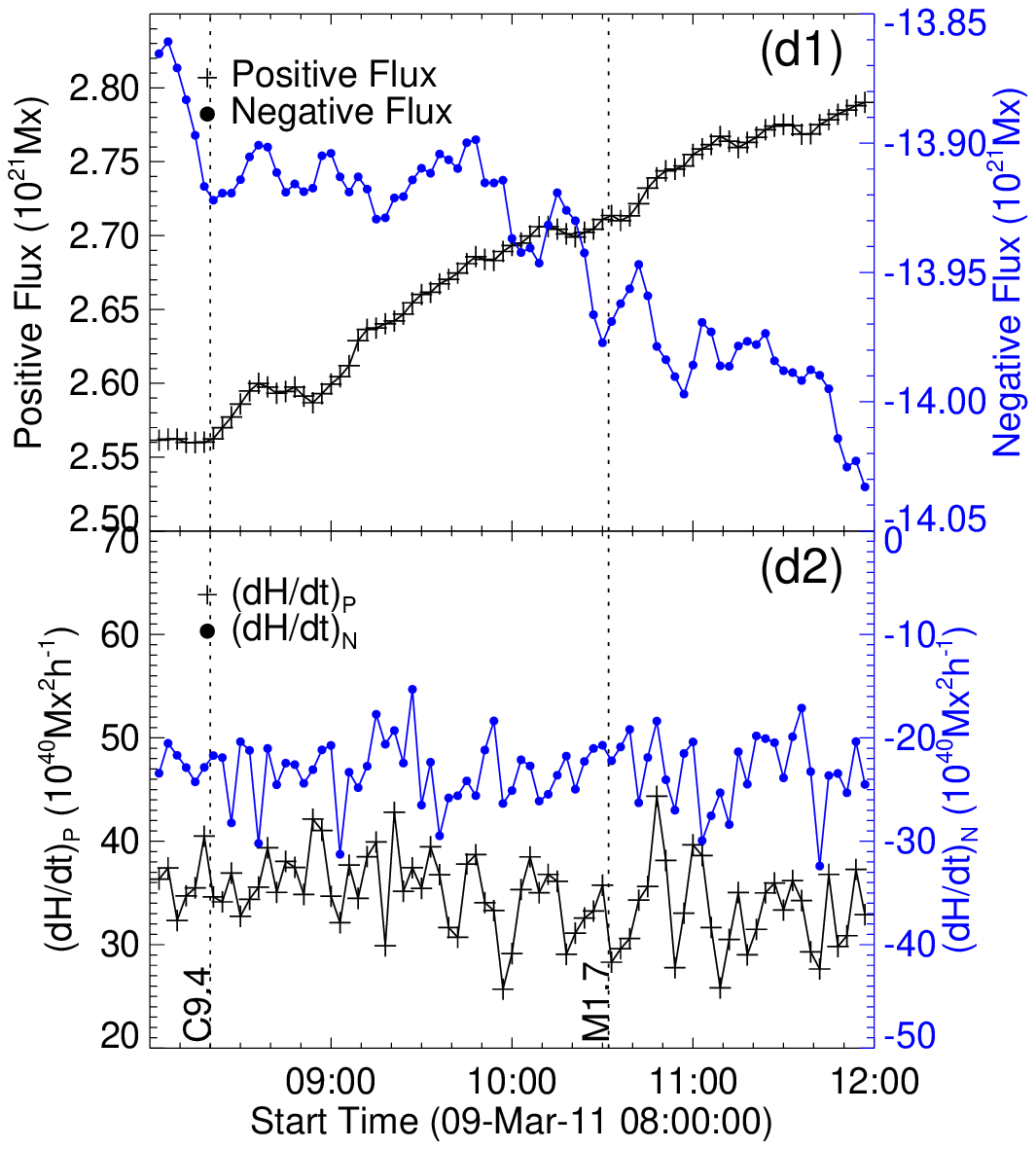}
\includegraphics[width=.325\textwidth,clip=,bb=66 365 368 708]{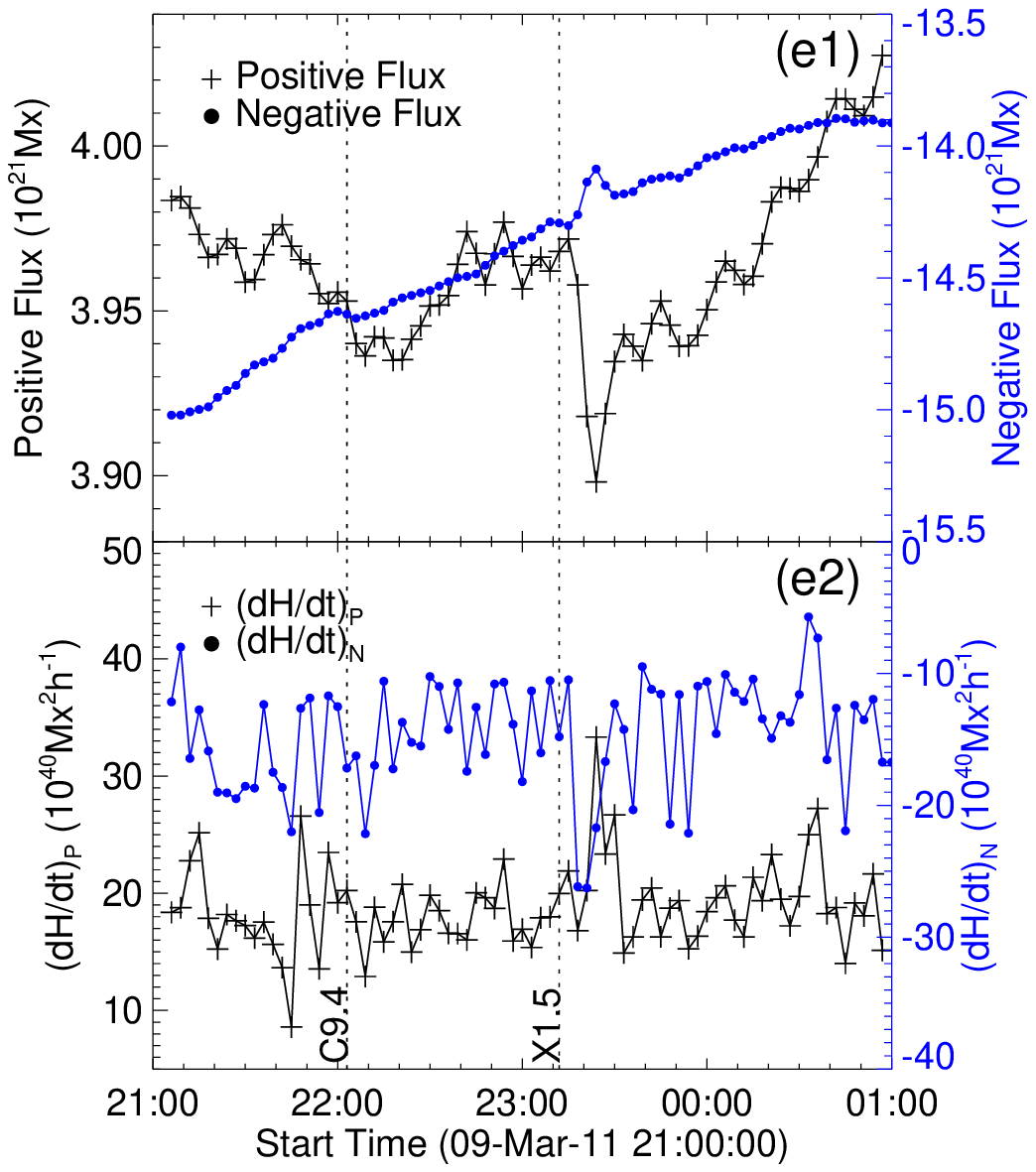}
\includegraphics[width=.325\textwidth,clip=,bb=66 365 368 708]{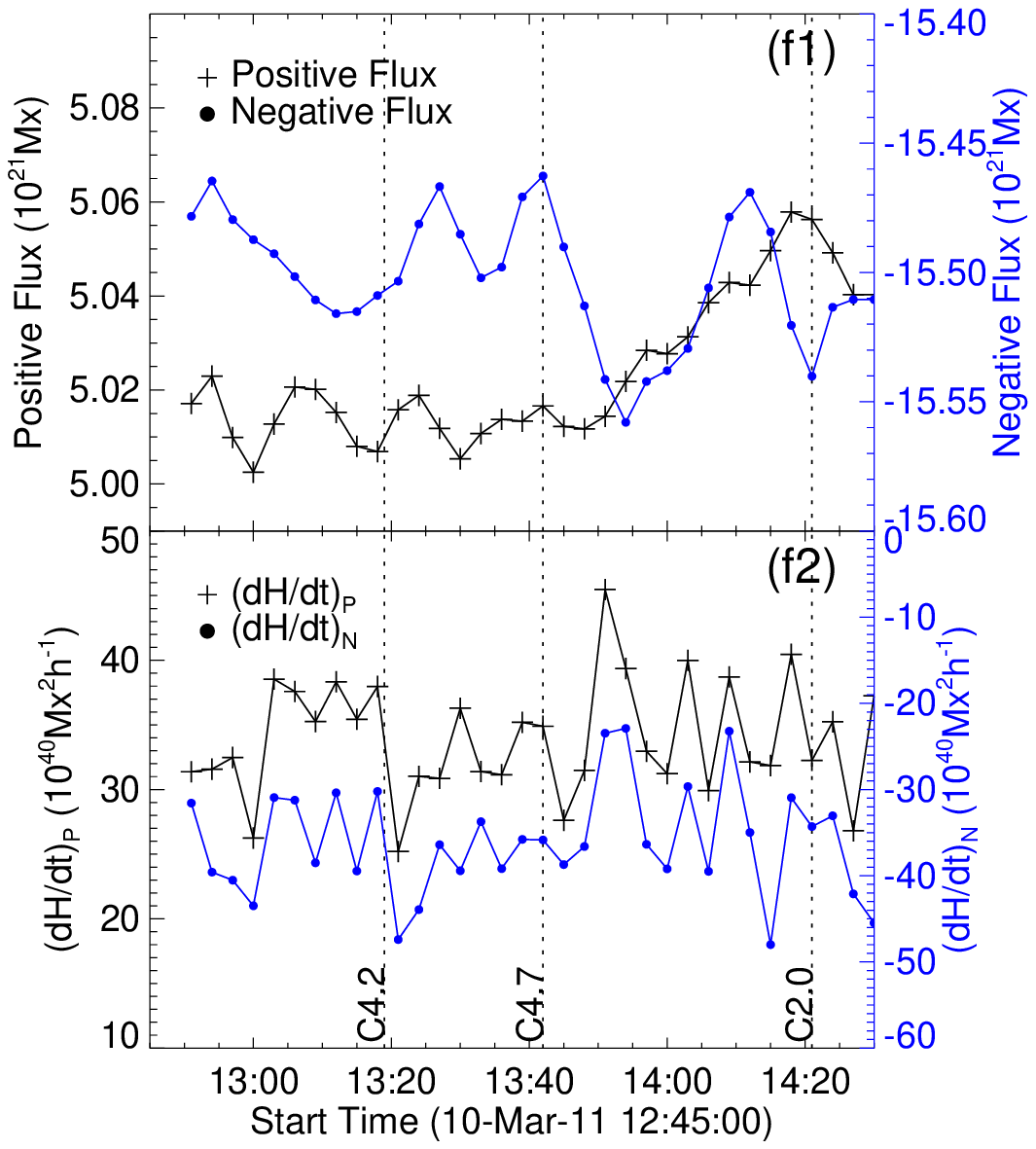}
\caption{Temporal profiles of magnetic and helicity fluxes during some selected
flare events in both ARs. Vertical dashed lines indicate onset time of flares
as labeled in each panel. See text for more details.} \label{PlotEve}
\end{figure*}

Similar mosaics of helicity flux distribution maps were made and examined  for
other events. The computed magnetic and helicity fluxes are plotted with time
in Figure~\ref{PlotEve}. The flare start time is shown in vertical dotted line
labeled with magnitude of the flare. It should be noted that we have not applied any
smoothing to the computed helicity rate signal in these panels. Magnetic fluxes
of both signs decreased abruptly with a dip during the impulsive period
following with injection of negative helicity flux in the dominant positive
helicity flux, during the M6.6, X2.2 flare events. Magnetic field measurements
could also be underestimated by 18-25\% due to enhanced core emission of
spectral line by the heating of the impulsive flare \citep{abramenko2004} as a
result of which the integrated flux profile could show such a dip during peak
phase of the flare. Interpretation of flux annihilation through reconnection
during this peak phase might be ambiguous due to this fact, although it could
be a possible consideration. In the post-flare phase, fluxes increased in both
polarities as field lines reorganized as a post-reconnection process. This
falls under the ``permanent'' real change related to the flare.

For smaller magnitude flares, transient effects may be  absent or not be
prominent in the impulsive phase. Therefore the measurements of magnetic fields
and the computed helicity rate signal are not expected to be affected during
the flare. Hence, they may indicate true transfer of helicity flux, except for
the theoretical uncertainties as mentioned above. In the case of the 14
February/13:47UT (C7.0) flare, shown in panels (b1)-(b2), indeed the variation
of helicity signal occurs without the variations in magnetic fluxes associated
to the flare-related effects. This may be an example of true transfer of
helicity of the flux system, but with the theoretical uncertainty in our
approach.

There are no significant variations in magnetic and helicity  fluxes
corresponding to the 09 March/09:23UT (C9.4), and 10:35UT (M1.7) flares. Large
amplitude fluctuations in both sign of helicity signals during the CME just
before the 09 March/22:03UT (C9.4) flare are apparent in panels (e1)-(e2). We
speculate that these fluctuations subsequently led to the initiation of the
prominent CME that followed the 09 March/23:13UT (X1.5) flare an hour later.
Similarly, the transient flare effects might be responsible for the abrupt
changes in magnetic fluxes resulting in variations of helicity injection signal
during the X1.5 flare (panels (e1)-(e2)). During the 10 March/13:19UT (C4.2),
13:42UT (C4.7) flares, the transfer of helicity flux from positive to negative,
negative to positive sign is clear from the panels (f1)-(f2), respectively.
These flares are of small magnitude, with no obvious flare-related artifacts.
Therefore, the observed helicity flux changes are expected to be true (with the
implicit theoretical uncertainty in the approach). A point to be noted is that
all large flares (M and X-class) may be involved with transient flare effects.
Therefore, it is better to look for helicity variations in small flares where
magnetic fields are expected to be less affected, making it easier to examine
the possible role of transfer of helicity flux. Thus, we consider the 14
February/13:47UT (C7.0), 10 March/13:19UT (C4.2) and 13:42UT (C4.7) flares to
be the best examples here, supporting the true transfer of helicity. It is not clear 
that whether the helicity transfer in these cases is related to permanent flare-effects.  

At present, it is difficult to say much about  the physical significance of
these variations over the AR in the corona, i.e., at the primary sites of the
flares. It would be particularly interesting to study the physical significance
of such injection along with the information of coronal connectivities ({\it
e.g.}, \citealt{chae2010}) as suggested by \citet{pariat2005} for understanding
the possible role of transfer of helicity flux during the flares/CMEs.

\subsection{Dependence of Helicity Injection Rate on the DAVE Parameters}
Computation of helicity injection rate involves the measurement of magnetic
field and the inferred horizontal velocities. Apart from the errors in the
measurements, the computations involving the DAVE method for deriving
velocities depend on two main parameters {\it viz.}, the time interval between
two successive magnetic maps, $\Delta$t, and the DAVE window size. For
obtaining optimized results, horizontal displacements of features during the
time interval $\Delta$t should be large enough to be well determined by DAVE.
Also, these displacements should be smaller than the selected window size. To
check our results for consistency, we carried out the DAVE calculations using
the time intervals $\Delta$t = 12, 24 and 36 minutes, while keeping the window
size fixed at $21 \times 18$ pixels. Then,  calculations were carried out for
different window sizes, {\it viz.}, 21$\times$18, 15$\times$12, 9$\times$6
while keeping $\Delta$t fixed at 36 minutes. Furthermore, to avoid the effect
arising from noise, we used a threshold of magnetic field at 10G, which is the
HMI precision. As the HMI provides 12 minute averaged data products, we
averaged them corresponding to our calculations at 24 (2 maps) and 36 (3 maps)
minutes.

\begin{figure*}
\centering
\includegraphics[width=.99\textwidth,clip=,bb=26 4 449 155]{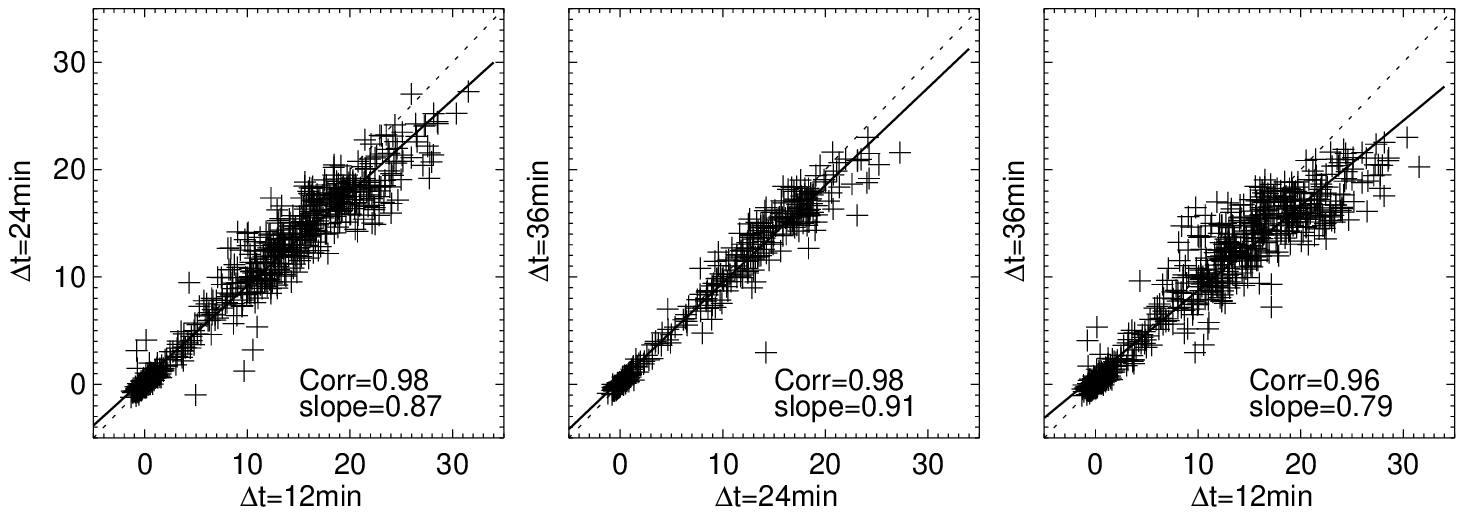}
\includegraphics[width=.99\textwidth,clip=,bb=26 5 449 155]{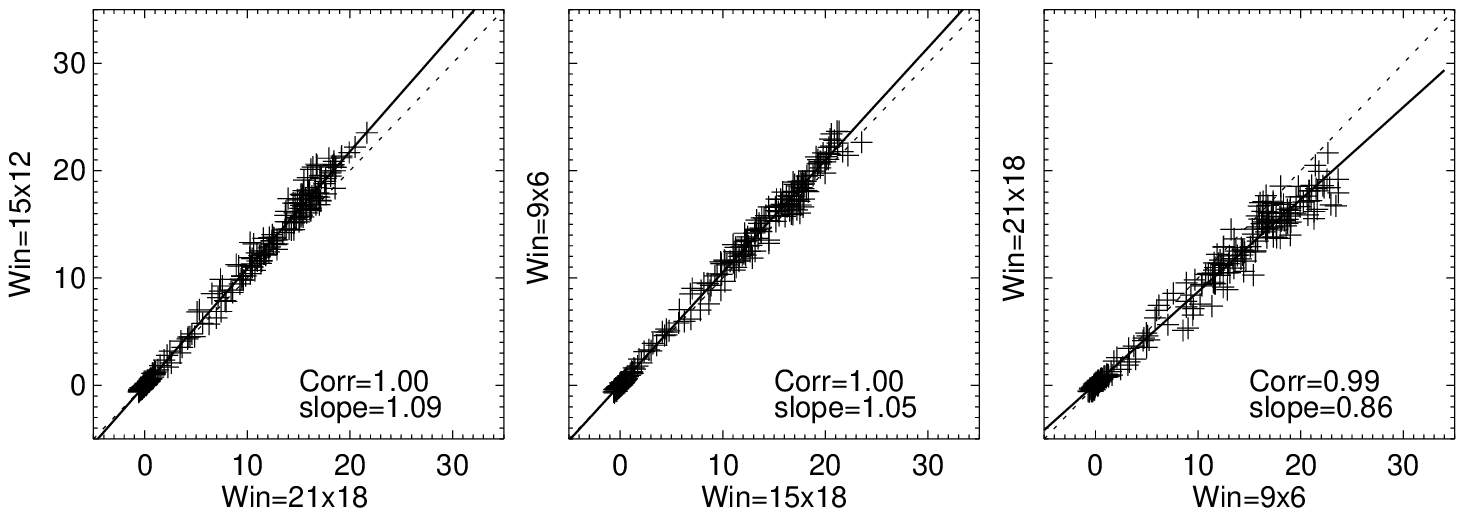}
\caption{Dependence of helicity injection rate (in units of 10$^{40}$ Mx$^2$h$^{-1}$) for
AR NOAA 11158 on (\emph{Top row}) the time interval $\Delta$t(minutes),  and
(\emph{bottom row}) the window size(pixel$^2$). The solid line represents the straight
line fit to the scattered data points whereas the dotted line indicates slope=1 line
for reference. Correlation coefficient and slope of the fitting are noted in each panel.} \label{depen_158}
\end{figure*}

The dependence of helicity injection rates on time interval $\Delta$t is shown in
Figure~\ref{depen_158}(top row) for NOAA 11158. The scattered data are fitted by straight
line in the least square sense. Due to the large volume of data, this computation is
tedious and time consuming. Therefore, results are shown here only for NOAA 11158,
but, we expect they are also valid for other ARs observed by the HMI. There
is an additional issue of unequally spaced data points to be addressed in case, for
example, we intend to plot the results for $\Delta$t=36 with $\Delta$t = 24 minutes. For
such cases, we used a cubic spline interpolation ({\it cf.}, \citealt{press1992}), to get
corresponding abscissa values for the ordinate points or vice-versa. Essentially, this
algorithm employs cubic polynomial between each pair of data points with the constraint
that the second and first derivatives of that polynomial are same at the end points
so that the resulting values are smooth. Table~\ref{TabHelDep} lists the minimum and
maximum values of helicity injection rates (dH/dt, in units of $10^{40}$Mx$^2$h$^{-1}$)
and the accumulated helicity ($\Delta H$, in units of $10^{42}$Mx$^2$) for the computational
runs carried out with various DAVE parameters as mentioned above.

\begin{table}[htbp]
  \centering
  \caption{Helicity injection rates and Accumulated  helicities at different DAVE parameters}
    \begin{tabular}{|cc|ccc|}
    \hline
    \multicolumn{2}{|c|}{DAVE parameters} & \multicolumn{3}{c|}{AR 11158} \\
    \hline
    $\Delta$t    & Window size & dH/dt    &    & $\Delta H$ \\
    min   & pixel$^2$ & min   & max   &  \\
    \hline
    12    & 21x18 & -18.98 & 31.54  & 14.16 \\
    24    & 21x18 & -7.48  & 27.27  & 13.09 \\
    36    & 21x18 & -1.06  & 22.52  & 12.96 \\
          &       &        &        &      \\
    36    & 21x18 & -1.06  & 22.52 & 12.96 \\
    36    & 15x12 & -1.06  & 25.02 & 13.51 \\
    36    & 9x6   & -1.28  & 26.8  & 14.22 \\
    \hline
    \end{tabular} \\
Units of dH/dt are $10^{40}$Mx$^2$h$^{-1}$ and $\Delta H$ are
$10^{42}$Mx$^2$
\label{TabHelDep}%
\end{table}%

It can be observed from the scatter plots that the helicity rates decreased
slightly as the time interval $\Delta$t is increased from 12 min to 36 min. The
fitted straight line deviates at a slope of 0.87 and 0.91 corresponding to
$\Delta t=12$ versus 24 and $\Delta t=24$ versus 36 min indicating that helicity
injection decreases by 13\% and 9\% respectively. This implies that short-lived
features and their dynamics have considerable contribution to helicity rates.
The helicity rates at intervals of 36min are lower by a factor of 21\% than that
at 12 min with worst correlation coefficient of 0.79. These effects in turn reflected
in the variation of accumulated helicity by 9\%. This implies that averaging in time
between 12-36 min has significant effect on injected helicity rates up to 13\%
corresponding to 9\% of variation in accumulated helicity.

The dependence of helicity injection rate on window size by keeping the time
interval $\Delta$t fixed at 36 minutes is shown in
Figure~\ref{depen_158}(bottom row). The slopes of 1.09 and 1.05 for the DAVE
windows $21\times18$ versus $15\times12$ and $15\times12$ versus $9\times6$
respectively, show increasing trend of helicity rates with decreasing
window size. Indeed, a scalable factor of 14\% reduction of helicity rate is
evident for windows $21\times18$ versus $9\times6$. Accumulated helicity also
showed this increased trend with decreased window size. A total variation of
$10\%$ is found, however, with the same trend of helicity injection rate
profiles which is discernible in correlation coefficient with the plots. A
maximum velocity of 1 km-s$^{-1}$ during the time interval of 12 min
corresponds to a plasma displacement of an arc-sec. Hence, for the window size
of 4.5\arcsec$\times$3\arcsec(9$\times$6 pixel$^2$), the issue of features
overflowing out of the window should not pose problem.

These results are consistent with those reported by \citet[their Figure
7]{chae2004}. They deduced and compared velocity and helicity rates by combinations of
time difference between magnetograms and LCT window size. Their rms velocity values
varied up to 0.6km/s at time interval of 5min. They found that smaller values of LCT
parameters result in larger amplitude fluctuations of the rate of helicity, with variation
within 10\%. We, in our computations, found maximum rms velocities for 12min, 24min and
36min in the AR as 0.95, 0.85 and 0.8km/s respectively. However for the window sizes
$21\times18$, $15\times12$ and $9\times6$, we obtained the rms velocities as 0.8, 0.9 and
1.5km/s respectively. These are higher by a factor of 2 compared to their values probably
due to the higher resolution and sensitivity of HMI as against the coarser spatial
resolution of MDI of 1.98\arcsec/pixel. Nevertheless, the variation in accumulated
helicity found in our analysis is within $10\%$; consistent with their result.

We thus, found the measured helicity injection rate to depend on the time
interval between the two successive magnetograms, i.e., the observational
cadence. The selected window size also influenced the measured quantities. Our
analysis suggests that it is better to use images averaged over up to 24
minutes with relatively small DAVE window size subjected to the overflow
condition as mentioned above. These are important considerations to derive
reasonable and meaningful results in addition to optimizing the computations
involving large data-sets.

\section{Discussions}
\label{disc}

Free energy storage and release are some of the most important problems in the
eruption physics of the Sun. There are essentially two effects that can supply
magnetic free energy and helicity from below the solar surface to the corona.
Flux emergence is the process in which vertical motions carry magnetic fluxes
through the photosphere. If the sub-surface fluxes emerging through the
photosphere are already twisted, then it will contribute to the injection of
helicity ({\it cf.}, the 1st term in Equation~\ref{eq_heli_ber}). Computation of this
term requires the knowledge of the vertical component of velocity and the
horizontal or transverse component of magnetic field. Flux motions in the form
of rotation or proper motions are another process that may efficiently supply
helicity injection ({\it cf.}, the 2nd term in Equation~\ref{eq_heli_ber}). The
helicity injected by solar differential rotation is rather small, less than
10\% of that contributed by the flux motions \citep{chae2004,demoulin2002b},
and has only a much longer term effect on helicity accumulation
\citep{devore2000}.

Magnetic helicity is a physical quantity having a positive or negative sign,
representing a right-handed or left-handed linkage of magnetic fluxes,
respectively. This means that if positive and negative helicities co-exist in a
single domain, magnetic reconnection can cancel magnetic helicity by merging
magnetic flux systems of opposite helicities. Helicity densities are not
gauge-invariant. It is only area-integrated relative helicity flux that is
gauge-invariant. In order to define true helicity flux density, the coronal
linkage needs to be provided \citep{pariat2005}, so the helicity flux density
inferred from tracking will not be precisely accurate. Our computations of
magnetic helicity injection in both ARs revealed that the distribution of
helicity flux is highly complicated in time and space. Even the sign of
helicity flux often changed within the AR.

It has been suggested earlier by several workers that magnetic helicity must play an
important role in flares as a substantial amount of helicity accumulation is found before
many events (\citealt{kusano1995, kusano1996, kusano2002}).  However, the correlation
between various magnetic field parameters and the flare index of an AR is not high
irrespective of the method used. This is an intrinsic problem for flare forecasting as
the occurrence of a flare depends not only on the amount of magnetic energy stored in an
AR, but also on how it is triggered. Thus, it appears that helicity accumulation might be
a necessary, but insufficient condition for the flares requiring a trigger even if a
magnetic system has enough non-potentiality. For instance, \citet{kusano2003}
suggested that coexistence of positive and negative helicities may be important for the onset
of flares.

Careful three-dimensional simulations have been carried out by \citet{linton2001} to
explore the physics of flux tube interaction for the co-helicity (same sign) or
counter-helicity (opposite sign). According to them, counter-helicity presented the most
energetic type of slingshot interaction in which flux is annihilated and twist is
canceled. In contrast, co-helicity exhibited very little interaction, and the flux
tubes bounced off resulting in negligible magnetic energy release.

Magnetic helicity in the solar corona is closely related to the photospheric
magnetic shear, which is usually defined as the extent of alignment of the
transverse component of magnetic field along the neutral or polarity inversion
line (PIL)\citep{ambastha1993}. Based on this idea, \cite{kusano2004} performed
a numerical simulation by applying a slow footpoint motion. This motion can reverse
the preloaded magnetic shear at the PIL resulting in a large scale eruption of
the magnetic arcade through a series of two different kinds of magnetic
reconnections. They proposed a model for solar flares in which magnetic
reconnection converts oppositely sheared field into shear-free fields.

We interpret our observations according to the above observational and
simulation aspects as follows. We have found flux interactions during the
X-class flares and associated CMEs as seen in Figure~\ref{58_vel} in the form
of continued shearing motion of SP2 around SN2 in AR 11158. Similar motions are
also associated with SP2 in AR 11166. In both ARs cases, the flare prone
regions (R2) had inhomogeneous the helicity flux distribution with mixed
helicities of both signs. Correspondingly, sudden impulsive peaks appeared in
the profiles of helicity injection due to the injection of negative signed
helicity during some flare events. These were also spatially correlated with
the observed flares. Opposite helicity flux tubes can interact easily leading
to reconnection, thereby unleashing explosive release of magnetic energy.
Impulsive variations of the magnetic helicity injection rate associated with
eruptive X- and M- class flares accompanied with CMEs were reported also by
\citet{moon2002}. Recently, \citet{park2010b} conjectured that the occurrence
of the X3.4 flare on 2006 December 13 was involved with the positive helicity
injection into an existing system of negative helicity. Further, a solar
eruption triggered by the interaction of two opposite-helicity flux
systems \citep{chandra2010,romano2011a}, and occurrence of flares in relation
to spatial distribution of helicity flux density \citep{romano2011b} were
reported. The main drawback of these findings is that the time span between two
magnetograms is more than the duration of the flare($\ge96$m), so the time rate of 
helicity could not be easily resolved at the onset time of the flare.
Therefore, our results appear to be consistent with the reports of
opposite helicity flux tubes reconnecting to trigger transient events.

However, it should be cautioned that we have not found such variations of
helicity flux clearly in all flare/CME events. From a quantitative analysis, we
found poor association of difference in helicity rate during flares to that of quiet times 
in AR NOAA 11158. This indicates such variations are not prominent or present during all
flares. Moreover, statistically significant association of such impulsive variations 
was found during CMEs compared to quiet times. There are many possible reasons for 
this poor association; one of them is time duration of helicity flux change. We first
interpolated the signal at 1 min interval from 12 min interval to get values as
required by the GOES flare times. Then, it was smoothed to a boxcar-averaging
window of 30 minutes to reduce fluctuations arising due to interpolation. Within 
start and stop times of flares, the averaged values of absolute variation were computed. 
Here, averaging might have diluted the original helicity variation, so comparison with the 
helicity variation during quiet times might not be valid. In any case, there is 
no better way to find appreciable variation in the helicity flux over background 
fluctuations to incorporate into the correlation analysis, unless individual events are
monitored manually to get variation timings. Despite these difficulties, 
statistically significant association of helicity flux is found during flares, but dominant 
association that is not statistically significant during CMEs in the AR 11166 by 
following the same approach.

Further, there are concerns about the flare-related effects on magnetic
field measurements resulting in misleading interpretation of helicity flux
transfer, in addition to the theoretical uncertainty with the assumption of
ideal magnetic field evolution in the approach. We therefore investigated this
issue using 3 min interval time sequence magnetograms. We found transient flare
effects resulting in spurious negative helicity flux distribution during the
X2.2, M6.6, and X1.5 flare events. Also, we indeed observed the true transfer of
helicity flux with variations of opposite sign helicity without such
flare-related effects in small flares such as the C7.0 on 14 February, C4.2 at
13:19UT, C4.7 at 13:42UT on 10 March. The important point to note is that we
found statistically significant association of helicity flux variations with
flares/CMEs in above cases of ARs at zero time lags. Also these variations are clear 
during the flare events (see Figure~\ref{PlotEve}) and not before their commencement. 
Therefore, it is difficult to suggest that these variations triggered the flares. A study
with the information of fieldline connectivity from coronal observations may be
expected to reveal the physical significance of the role of helicity transfer
during these events.

Our computed helicity rates involving photospheric flux motions include the flux
emergence term as explained by \citet{demoulin2003}. By a simple geometrical argument,
horizontal foot-point velocity ($\mathbf{u}$, here the DAVE velocity) can be written
in terms of horizontal and vertical plasma velocities, $\mathbf{v}_{h}$, $v_{n}$,
respectively:

\begin{equation}
\mathbf{u}=\mathbf{{v}_{h}}-\frac{v_n}{B_n}\mathbf{B_h}.
\end{equation}

\begin{figure*}
\centering
\includegraphics[width=.97\textwidth,clip=,bb=21 2 355 196]{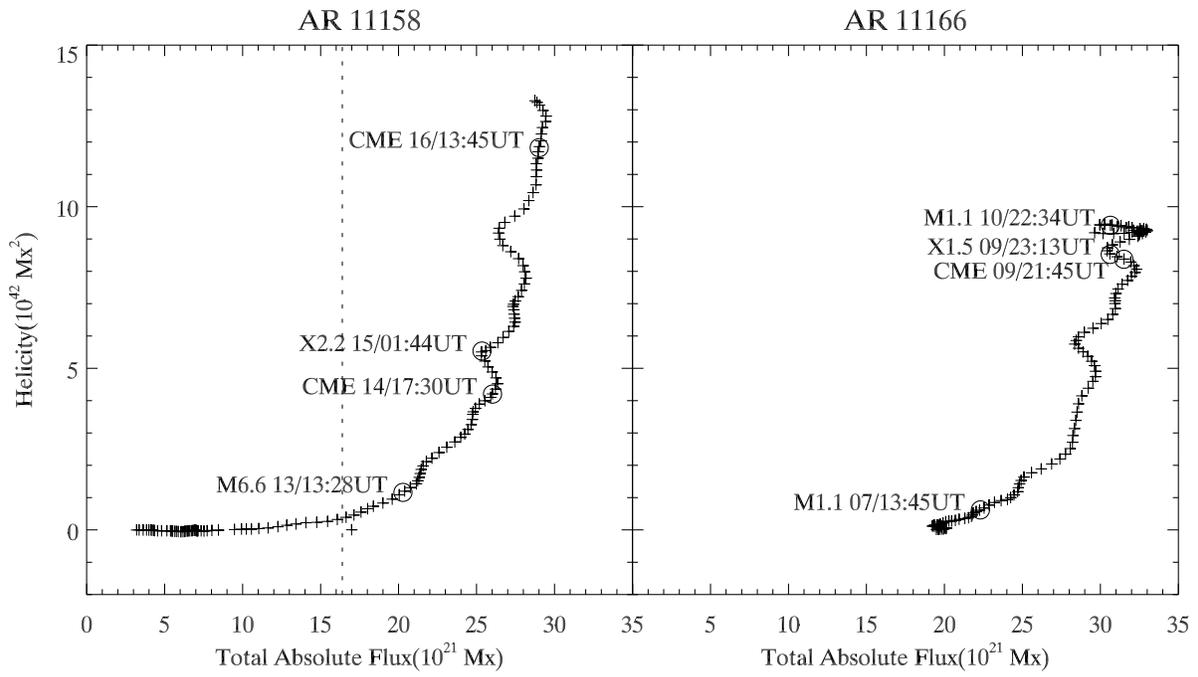}
\caption{Plot of accumulated helicity with total absolute flux computed for NOAA
11158(\emph{Left}) and NOAA 11166(\emph{Right}). The flare/CME events are labeled and shown by
circles in each panel.} \label{flx_Vs_hel}
\end{figure*}

\noindent From this relation, it is not possible to infer as to which term,
{\it viz.,} the flux emergence or flux motions, governs the level of activity of the ARs.
To resolve this difficulty, we have plotted the integrated absolute flux and accumulated
helicity computed over the ARs, as shown in Figure~\ref{flx_Vs_hel}.

Evidently, the accumulated helicity increased monotonically with the emergence
of magnetic flux in the AR in its first phase (marked by the vertical dashed
line for NOAA 11158). After this phase followed the next, the active phase,
where an appreciable increase of helicity occurred with only small variation in
the flux, i.e., where little emergence of fluxes occurred. This rapid increase
in helicity in the second phase could be interpreted as the dominant
contribution of the flux motions. Intermittent mass expulsions in the form of
CMEs transferred away the excess helicity. The extent of this transfer,
however, is not clear from this plot, although one can make plausible
conclusions from the timings of the flares and CMEs. The X-class flares with
associated CMEs in both ARs occurred at a slowing phase of helicity
accumulation by negative helicity injection. These facts add to the cases as
reported by \citet{park2010a}.

Moreover, it can be inferred for AR 11158, that less than 25\% of the
total helicity flux accumulated with the emergence of the first 75\% of the
magnetic flux. Most of the helicity flux (from about $3-13\times10^{42}$Mx$^2$)
was accompanied by very little flux emergence (about $3\times10^{21}$Mx out of
the $30\times10^{21}$Mx). Therefore, more than 75\% of the helicity flux came
with only 10\% of the total magnetic flux. Similarly, the first 60\%
($19.5-28.0\times10^{21}$Mx) of total magnetic flux was associated to less than
30\% ($3\times10^{42}$Mx$^2$ of $9.5\times10^{42}$Mx$^2$) of the total helicity 
flux in AR 11166. This implies that more than 70\% of total helicity flux was 
accompanied with less than 40\% of total magnetic flux.
These two cases are thus contrary to the findings of \citet{jeong2007} stating
that most of the helicity flux occurs during flux emergence. Our study suggests
that flux emergence may not always play a major role in
accumulating helicity flux. It is also evident that although flux
emergence is necessary but horizontal motions also played crucial and
dominant role over emergence term in increasing the complexity of magnetic
structures contributing to the helicity flux. Therefore, we suggest that the
horizontal flux motions contributed further, in addition to the emergence term,
in creating more complex magnetic structures that caused the observed eruptive
phenomena.

\section{Summary}
\label{Sum}
We have studied the evolution of magnetic fluxes, horizontal flux motions,
helicity injection and their relationship with the eruptive transient events in
two recent flare (CME) productive ARs, NOAA 11158 and NOAA 11166 of 2011
February and March, respectively. We have used high resolution, high cadence
data provided by SDO-HMI for these ARs which were in their emerging and active
phases. The emerging AR consisted of rotating sunspots with increasing flux
indicating emergence of twisted flux from the sub-photospheric layers.
This indicated the transfer of twist or helicity injection through the
photosphere to the outer atmosphere.

We suggest that strong shear motions that include rotational and proper motions played
significant role in  most of the events in addition to the flux emergence. Such
motions are crucial in twisting or shearing the magnetic field lines and for
further flux interactions. AR NOAA 11158 consisted of a CME-prone site of
rotating main sunspot along with emerging flux of opposite sign and moving magnetic
feature. It also had a flare-prone site consisting of self-rotating sunspot(SP2)
moving about a sunspot of opposite sign(SN2), leading to flux interaction. These
motions are likely to form the sigmoidal structures, which are unstable, and  more likely
to produce eruptive events. A huge expulsion as CME on 2011 February 14/17:30UT occurred
in the former site and a white light, energetic X2.2 flare on 2011 February 15/01:44UT
occurred in the later site. The other case, AR NOAA 11166 was already in its active phase
with further increasing content of flux as it evolved. Group motions of diffused fluxes
merging to form a bigger sunspot manifested major shear motions in addition to outward flows
from sunspot. A large CME on 2011 March 09/21:45UT, followed by an X1.5 flare, was one of
the major events in this AR.

AR NOAA 11158 injected $14.16\times10^{42}$Mx$^2$ while AR NOAA 11166 injected
$9.5\times10^{42}$Mx$^2$ helicity during the six days' period of their
evolution. These are consistent with the previously reported order of helicity
accumulation \cite[{\it e.g.},][]{park2010a}. It appears that due to the
presence of rotational motions, the former AR accumulated larger amount of
helicity accounting for its greater activity in the form of flares and CMEs. It
is also evident that flux emergence is necessary and their motions are crucial
in additionally accounting for the accumulated amount of helicity to the
emergence term. In both ARs, X-class flares with associated CMEs were observed
in the decreasing phase of helicity accumulation by the injection of opposite
helicity.

Apart from the instrumental and computational errors, the estimation of
helicity injection rates are also affected by the choice of DAVE parameters
used to track the motion of the fluxes. Helicity injection rates are found to
decrease up to 13\% by increasing the time interval between magnetograms from
12 to 36 min whereas an increasing trend upto 9\% resulted by decreasing the
window size from $21\times18$ to $9\times6$ pixel$^2$, with a total variation
of 10\% in the deduced value of accumulated helicity.

The time profile of helicity rate exhibited sudden sharp variations during
some flare events due to injection of opposite helicity flux into the existing
system of helicity flux. In both ARs, the flare prone regions (R2) had 
inhomogeneous helicity flux distribution with mixed helicities of both signs and that of 
CME prone regions had almost homogeneous distribution of helicity flux dominated by single sign. 
A quantitative analysis was carried out to show the association of these variations to 
the timings of flares/CMEs. For the AR 11158, we find a marginally significant 
association of helicity flux with CMEs but not flares, while for the AR 11166, we find 
marginally significant association of helicity flux with flares but not CMEs. Moreover, 
these variations of helicity flux may not reflect true transfer; there exists flare-related 
transient effects and theoretical uncertainties resulting to these variations. We believe 
the helicity transfer in the cases of C7.0 on 14 February, C4.2 at 13:19UT, C4.7 at 13:42UT on 10 
March to be true, without flare-related transient effect but with theoretical
uncertainty in the approach.

Therefore, to further strengthen the above evidences of true helicity transfer,
it would be worthwhile to scrutinize more flare/CMEs cases using 3 min cadence
magnetic observations, over a period of a day or so. This will enable one to
find detectable changes in helicity flux signal during smaller magnitude flares
with less transient-flare effects. Interpreting the physical significance of
such variations using the information of coronal connectivities will be another
important aspect to add further to the present knowledge of helicity physics.
Our study reveals that the spatial information of helicity injection is a key
factor to understand its role in the flares/CMEs.


\acknowledgements The data have been used here courtesy of NASA/SDO and HMI
science team. We thank Dr. Etienne Pariat for checking our helicity program
with comments and suggestions. The author expresses his gratitude to Prof. P.
Venkatakrishnan for some useful discussions on the concept of helicity. We
thank an anonymous referee for carefully reading the manuscript and making
valuable comments which led to improved clarity and readability of the
manuscript.

We thank Mr.Jigar Raval and Mr.Anish Parwage for their help in running program on
one of the nodes of 3TFLOP HPC cluster at PRL computer center.

\bibliographystyle{apj}

\begin{thebibliography}{52}
\expandafter\ifx\csname natexlab\endcsname\relax\def\natexlab#1{#1}\fi

\bibitem[{{Abramenko} \& {Baranovsky}(2004)}]{abramenko2004}
{Abramenko}, V.~I., \& {Baranovsky}, E.~A. 2004, \solphys, 220, 81

\bibitem[{{Ambastha} {et~al.}(1993){Ambastha}, {Hagyard}, \&
  {West}}]{ambastha1993}
{Ambastha}, A., {Hagyard}, M.~J., \& {West}, E.~A. 1993, \solphys, 148, 277

\bibitem[{{Berger} \& {Field}(1984)}]{berger1984}
{Berger}, M.~A., \& {Field}, G.~B. 1984, \jfm, 147, 133

\bibitem[{{Canfield} {et~al.}(1999){Canfield}, {Hudson}, \&
  {McKenzie}}]{canfield1999}
{Canfield}, R.~C., {Hudson}, H.~S., \& {McKenzie}, D.~E. 1999, \grl, 26, 627

\bibitem[{{Chae}(2001)}]{chae2001a}
{Chae}, J. 2001, \apjl, 560, L95

\bibitem[{{Chae}(2007)}]{chae2007}
---. 2007, Advances in Space Research, 39, 1700

\bibitem[{{Chae} {et~al.}(2010){Chae}, {Goode}, {Ahn}, {Yurchysyn},
  {Abramenko}, {Andic}, {Cao}, \& {Park}}]{chae2010}
{Chae}, J., {Goode}, P.~R., {Ahn}, K., {Yurchysyn}, V., {Abramenko}, V.,
  {Andic}, A., {Cao}, W., \& {Park}, Y.~D. 2010, \apjl, 713, L6

\bibitem[{{Chae} {et~al.}(2004){Chae}, {Moon}, \& {Park}}]{chae2004}
{Chae}, J., {Moon}, Y.-J., \& {Park}, Y.-D. 2004, \solphys, 223, 39

\bibitem[{{Chae} {et~al.}(2001){Chae}, {Wang}, {Qiu}, {Goode}, {Strous}, \&
  {Yun}}]{chae2001b}
{Chae}, J., {Wang}, H., {Qiu}, J., {Goode}, P.~R., {Strous}, L., \& {Yun},
  H.~S. 2001, \apj, 560, 476

\bibitem[{{Chandra} {et~al.}(2010){Chandra}, {Pariat}, {Schmieder}, {Mandrini},
  \& {Uddin}}]{chandra2010}
{Chandra}, R., {Pariat}, E., {Schmieder}, B., {Mandrini}, C.~H., \& {Uddin}, W.
  2010, \solphys, 261, 127

\bibitem[{{D{\'e}moulin} \& {Berger}(2003)}]{demoulin2003}
{D{\'e}moulin}, P., \& {Berger}, M.~A. 2003, \solphys, 215, 203

\bibitem[{{D{\'e}moulin} {et~al.}(2002){D{\'e}moulin}, {Mandrini}, {van
  Driel-Gesztelyi}, \& {et al}}]{demoulin2002b}
{D{\'e}moulin}, P., {Mandrini}, C.~H., {van Driel-Gesztelyi}, L., \& {et al}.
  2002, \aap, 382, 650

\bibitem[{{DeVore}(2000)}]{devore2000}
{DeVore}, C.~R. 2000, \apj, 539, 944

\bibitem[{{Finn} \& {Antonsen}(1985)}]{finn1985}
{Finn}, J.~M., \& {Antonsen}, T.~M. 1985, \cpcf, 9, 111

\bibitem[{{Hagyard} \& {Pevtsov}(1999)}]{hagyard1999}
{Hagyard}, M.~J., \& {Pevtsov}, A.~A. 1999, \solphys, 189, 25

\bibitem[{{Howard} {et~al.}(1990){Howard}, {Harvey}, \& {Forgach}}]{howard1990}
{Howard}, R.~F., {Harvey}, J.~W., \& {Forgach}, S. 1990, \solphys, 130, 295

\bibitem[{{Jeong} \& {Chae}(2007)}]{jeong2007}
{Jeong}, H., \& {Chae}, J. 2007, \apj, 671, 1022

\bibitem[{{Kusano} {et~al.}(2002){Kusano}, {Maeshiro}, {Yokoyama}, \&
  {Sakurai}}]{kusano2002}
{Kusano}, K., {Maeshiro}, T., {Yokoyama}, T., \& {Sakurai}, T. 2002, \apj, 577,
  501

\bibitem[{{Kusano} {et~al.}(2004){Kusano}, {Maeshiro}, {Yokoyama}, \&
  {Sakurai}}]{kusano2004}
---. 2004, \apj, 610, 537

\bibitem[{{Kusano} \& {Nishikawa}(1996)}]{kusano1996}
{Kusano}, K., \& {Nishikawa}, K. 1996, \apj, 461, 415

\bibitem[{{Kusano} {et~al.}(1995){Kusano}, {Suzuki}, \&
  {Nishikawa}}]{kusano1995}
{Kusano}, K., {Suzuki}, Y., \& {Nishikawa}, K. 1995, \apj, 441, 942

\bibitem[{{Kusano} {et~al.}(2003){Kusano}, {Yokoyama}, {Maeshiro}, \&
  {Sakurai}}]{kusano2003}
{Kusano}, K., {Yokoyama}, T., {Maeshiro}, T., \& {Sakurai}, T. 2003, Advances
  in Space Research, 32, 1931

\bibitem[{{LaBonte} {et~al.}(2007){LaBonte}, {Georgoulis}, \&
  {Rust}}]{labonte2007}
{LaBonte}, B.~J., {Georgoulis}, M.~K., \& {Rust}, D.~M. 2007, \apj, 671, 955

\bibitem[{{Leka} {et~al.}(1996){Leka}, {Canfield}, {McClymont}, \& {van
  Driel-Gesztelyi}}]{leka1996}
{Leka}, K.~D., {Canfield}, R.~C., {McClymont}, A.~N., \& {van Driel-Gesztelyi},
  L. 1996, \apj, 462, 547

\bibitem[{{Linton} {et~al.}(2001){Linton}, {Dahlburg}, \&
  {Antiochos}}]{linton2001}
{Linton}, M.~G., {Dahlburg}, R.~B., \& {Antiochos}, S.~K. 2001, \apj, 553, 905

\bibitem[{{Longcope}(2004)}]{longcope2004}
{Longcope}, D.~W. 2004, \apj, 612, 1181

\bibitem[{{Maurya} \& {Ambastha}(2009)}]{maurya2009}
{Maurya}, R.~A., \& {Ambastha}, A. 2009, \solphys, 258, 31

\bibitem[{{Maurya} {et~al.}(2012){Maurya}, {Vemareddy}, \&
  {Ambastha}}]{maurya2012}
{Maurya}, R.~A., {Vemareddy}, P., \& {Ambastha}, A. 2012, \apj, 747, 134

\bibitem[{{Moon} {et~al.}(2002){Moon}, {Chae}, {Wang}, {Choe}, \&
  {Park}}]{moon2002}
{Moon}, Y.~J., {Chae}, J., {Wang}, H., {Choe}, G.~S., \& {Park}, Y.~D. 2002,
  \apj, 580, 528

\bibitem[{{Nindos} {et~al.}(2003){Nindos}, {Zhang}, \& {Zhang}}]{nindos2003}
{Nindos}, A., {Zhang}, J., \& {Zhang}, H. 2003, \apj, 594, 1033

\bibitem[{{November} \& {Simon}(1988)}]{november1988}
{November}, L.~J., \& {Simon}, G.~W. 1988, \apj, 333, 427

\bibitem[{{Pariat} {et~al.}(2005){Pariat}, {D{\'e}moulin}, \&
  {Berger}}]{pariat2005}
{Pariat}, E., {D{\'e}moulin}, P., \& {Berger}, M.~A. 2005, \aap, 439, 1191

\bibitem[{{Park} {et~al.}(2010{\natexlab{a}}){Park}, {Chae}, {Jing}, {Tan}, \&
  {Wang}}]{park2010b}
{Park}, S.-H., {Chae}, J., {Jing}, J., {Tan}, C., \& {Wang}, H.
  2010{\natexlab{a}}, \apj, 720, 1102

\bibitem[{{Park} {et~al.}(2010{\natexlab{b}}){Park}, {Chae}, \&
  {Wang}}]{park2010a}
{Park}, S.-h., {Chae}, J., \& {Wang}, H. 2010{\natexlab{b}}, \apj, 718, 43

\bibitem[{{Petrie} \& {Sudol}(2010)}]{petrie2010}
{Petrie}, G.~J.~D., \& {Sudol}, J.~J. 2010, \apj, 724, 1218

\bibitem[{{Pevtsov} {et~al.}(1995){Pevtsov}, {Canfield}, \&
  {Metcalf}}]{pevtsov1995}
{Pevtsov}, A.~A., {Canfield}, R.~C., \& {Metcalf}, T.~R. 1995, \apjl, 440, L109

\bibitem[{{Pevtsov} {et~al.}(2008){Pevtsov}, {Canfield}, {Sakurai}, \&
  {Hagino}}]{pevtsov2008}
{Pevtsov}, A.~A., {Canfield}, R.~C., {Sakurai}, T., \& {Hagino}, M. 2008, \apj,
  677, 719

\bibitem[{{Press} {et~al.}(1992){Press}, {Teukolsky}, {Vetterling}, \&
  {Flannery}}]{press1992}
{Press}, W.~H., {Teukolsky}, S.~A., {Vetterling}, W.~T., \& {Flannery}, B.~P.
  1992, {Numerical recipes in C. The art of scientific computing}, ed. {Press,
  W.~H., Teukolsky, S.~A., Vetterling, W.~T., \& Flannery, B.~P. }

\bibitem[{{Qiu} \& {Gary}(2003)}]{qiu2003}
{Qiu}, J., \& {Gary}, D.~E. 2003, \apj, 599, 615

\bibitem[{{Romano} {et~al.}(2011){Romano}, {Pariat}, {Sicari}, \&
  {Zuccarello}}]{romano2011a}
{Romano}, P., {Pariat}, E., {Sicari}, M., \& {Zuccarello}, F. 2011, \aap, 525,
  A13

\bibitem[{{Romano} \& {Zuccarello}(2011)}]{romano2011b}
{Romano}, P., \& {Zuccarello}, F. 2011, \aap, 535, A1

\bibitem[{{Schou} {et~al.}(2012){Schou}, {Scherrer}, {Bush}, {Wachter}, \& {et
  al}}]{schou2012}
{Schou}, J., {Scherrer}, P.~H., {Bush}, R.~I., {Wachter}, R., \& {et al}. 2012,
  \solphys, 275, 229

\bibitem[{{Schuck}(2005)}]{schuck2005}
{Schuck}, P.~W. 2005, \apjl, 632, L53

\bibitem[{{Schuck}(2006)}]{schuck2006}
---. 2006, \apj, 646, 1358

\bibitem[{{Schuck}(2008)}]{schuck2008}
---. 2008, \apj, 683, 1134

\bibitem[{{Sudol} \& {Harvey}(2005)}]{sudol2005}
{Sudol}, J.~J., \& {Harvey}, J.~W. 2005, \apj, 635, 647

\bibitem[{{Tian} {et~al.}(2011){Tian}, {D{\'e}moulin}, {Alexander}, \&
  {Zhu}}]{tian2011}
{Tian}, L., {D{\'e}moulin}, P., {Alexander}, D., \& {Zhu}, C. 2011, \apj, 727,
  28

\bibitem[{{Tiwari} {et~al.}(2009){Tiwari}, {Venkatakrishnan}, {Gosain}, \&
  {Joshi}}]{sanjiv2009}
{Tiwari}, S.~K., {Venkatakrishnan}, P., {Gosain}, S., \& {Joshi}, J. 2009,
  \apj, 700, 199

\bibitem[{{Venkatakrishnan} \& {Gary}(1989)}]{venkat1989}
{Venkatakrishnan}, P., \& {Gary}, G.~A. 1989, \solphys, 120, 235

\bibitem[{{Wang} \& {Liu}(2010)}]{wang2010}
{Wang}, H., \& {Liu}, C. 2010, \apjl, 716, L195

\bibitem[{{Welsch} {et~al.}(2007){Welsch}, {Abbett}, {De Rosa}, {Fisher}, \&
  {et al}}]{welsh2007}
{Welsch}, B.~T., {Abbett}, W.~P., {De Rosa}, M.~L., {Fisher}, G.~H., \& {et
  al}. 2007, \apj, 670, 1434

\bibitem[{{Welsch} {et~al.}(2004){Welsch}, {Fisher}, {Abbett}, \&
  {Regnier}}]{welsh2004}
{Welsch}, B.~T., {Fisher}, G.~H., {Abbett}, W.~P., \& {Regnier}, S. 2004, \apj,
  610, 1148

\end{thebibliography}


\end{document}